# Modeling False Data Injection Attacks in Integrated Electricity-Gas Systems

Rong-Peng Liu, Xiaozhe Wang, Zuyi Li, and Rawad Zgheib

***Abstract*—Integrated electricity-gas systems (IEGSs) rely heavily on communication systems and are vulnerable to cyberattacks. In order to gain insights into intruders' behavior and design tailored detection and mitigation methods, this paper studies the modeling of false data injection attacks (FDIAs) on IEGSs. First, we design a tailored static state estimation model and a bad data detection method for IEGSs. Then, we develop FDIAs on IEGSs with complete network topology and parameter information and give conditions for ensuring the stealthiness of these FDIAs. Particularly, the FDIAs consider the cyberattack interdependency caused by power-gas coupling facilities. Next, we develop FDIAs on IEGSs when intruders have only local network topology and parameter information of an IEGS. At last, we explore FDIAs on IEGSs when intruders have only local network topology information of an IEGS and mathematically prove the existence of FDIAs, specifically targeting gas compressors. Simulation results validate the effectiveness of the proposed FDIAs on IEGSs with both complete and incomplete network information.**

## NOMENCLATURE

$\mathcal{P}_d/\mathcal{G}_d$ Set of power/gas loads.

$\mathcal{P}_g/\mathcal{G}_g/\mathcal{G}_w$ Set of coal-fired generators/gas-fired generators/gas wells.

$\mathcal{P}_l/\mathcal{G}_l/\mathcal{G}_c$ Set of power transmission lines/gas passive pipelines/gas compressors (gas active pipelines).

$\mathcal{P}_n/\mathcal{G}_n/\mathcal{P}'_n$ Set of power buses/gas nodes/power buses equipped with phasor measurement units (PMUs).

$\mathcal{P}_f$ Set of power-to-gas (P2G) facilities.

$\mathrm{C}_{ij}^{\max}$ Transmission limit of gas compressor $\{i, j\}$.

$\mathrm{G}_i^{\min}/\mathrm{G}_i^{\max}$ Minimum/maximum gas injection at gas node $i$.

$\mathrm{G}_{ij}/\mathrm{B}_{ij}$ Conductance/susceptance of power transmission line $\{i, j\}$.

$\mathrm{P}_d/\mathrm{Q}_d/\mathrm{G}_d$ Real power/reactive power/gas load $d$.

$\mathrm{P}_i^{\min}/\mathrm{P}_i^{\max}$ Minimum/maximum real power injection at power bus $i$.

$\mathrm{Q}_i^{\min}/\mathrm{Q}_i^{\max}$ Minimum/maximum reactive power injection at power bus $i$.

$\mathrm{V}_i^{\min}/\mathrm{V}_i^{\max}$ Minimum/maximum voltage magnitude at power bus $i$.

$\mathrm{S}_{ij}^{\max}/\mathrm{G}_{ij}^{\max}$ Transmission limit of power transmission line $\{i, j\}$/gas passive pipeline $\{i, j\}$.

$\mathrm{W}_{ij}$ Weymouth constant of gas passive pipeline$\{i, j\}$.

$\alpha_{ij}/\gamma_g/\gamma_f$ Compression ratio of gas compressor $\{i, j\}$/electricity-gas conversion ratio of gas-fired power generator $g$/P2G facility $f$.

$\Pi_i^{\min}/\Pi_i^{\max}$ Minimum/maximum nodal pressure at gas node $i$.

$\theta_i^{\max}$ Maximum phase angle at power bus $i$.

$g_{ij}/c_{ij}$ Gas flow in gas passive pipeline/gas compressor $ij$.

$g_i/g_w$ Gas injection at gas node $i$/gas output of gas well $w$.

$g_f$ Gas output/energy consumption rate of P2G facility $f$.

$p_g/q_g$ Real/reactive power output of power generator $g$.

$p_{ij}/q_{ij}$ Real/reactive power flow in power transmission line $\{i, j\}$.

$p_i/q_i$ Real/reactive power injection at power bus $i$.

$v_i/\pi_i$ Voltage magnitude/nodal pressure at power bus/gas node $i$.

$\theta_i$ Phase angle at power bus $i$.

$\Delta\bullet$ Injected false data in the measurement of $\bullet$, where $\bullet \in \{g_{ij}, c_{ij}, g_i, p_{ij}, q_{ij}, p_i, q_i, v_i, \pi_i, \theta_i\}$. For example, $\Delta g_{ij}$ denotes the injected false data in the gas flow measurement in gas passive pipeline $\{i, j\}$.

R. Liu and X. Wang are with the Department of Electrical and Computer Engineering, McGill University, Montreal, QC H3A 0E9, Canada (e-mail: rpliu@eee.hku.hk/rongpeng.liu@mail.mcgill.ca; xiaozhe.wang2@mcgill.ca).

Z. Li is with the College of Electrical Engineering and Polytechnic Institute, Zhejiang University, Hangzhou 310027, China (email: lizuyi@zju.edu.cn).

R. Zgheib is with Hydro-Quebec Research Institute (IREQ) (email: zgheib.rawad2@hydroquebec.com).

## I. INTRODUCTION

Integrated electricity-gas systems (IEGSs) have been widely constructed to facilitate cost-effective gas-fired power generation [1]. However, their reliance on communication systems (to maintain synchronous operation of power and gas subsystems in an IEGS) makes them vulnerable to cyberattacks. Recently, the soaring number of cyber incidents in the energy sector [2] and the severe consequences [3], [4] highlight the urgent need to ensure IEGS cybersecurity.

Prior research extensively studied cyberattack paradigms on *power systems* [5]-[7]. Reference [5] pointed out the vulnerability of power systems to false data injection attacks (FDIAs), a kind of cyberattack that can bypass bad data detection (BDD) and compromise direct current (DC) state estimation (SE). Reference [6] analyzed the impact of FDIAs on alternating current (AC) SE. Reference [7] designed sequential FDIAs on battery energy storage systems in distribution systems. These studies promote follow-up research on the cybersecurity of the power subsector and other energy subsectors.

Recently, cyberattacks on *IEGSs* have intrigued researchers' interests [8]-[15]. Reference [8] developed FDIAs on the gas system in an IEGS and evaluated their impact on the interconnected power system. Reference [9] investigated FDIAs involving false gas supply information for gas-fired generators in IEGSs. Reference [10] explored cyberattacks on compressor stations in IEGSs. Reference [11] proposed FDIAs on power and gas pipeline monitoring systems in an IEGS, separately, for disabling gas-fired generators and disconnecting entire IEGSs. References [12]-[15] developed load redistribution attacks on the distribution system in IEGSs, load redistribution attacks on IEGSs, FDIAs on IEGSs, and FDIAs on integrated electricity-gas-water systems, respectively. These works [8]-[15] contribute to identifying IEGS cyber-vulnerability in practice. Despite significant advancements, distinct gaps still exist in the study of *FDIAs on IEGSs* in [8]-[15]: i) the FDIAs on IEGSs overlook critical cyberattack interdependency (CAI[#a]), including the interdependencies between power and gas systems through, for example, gas-fired power generators and power-to-gas (P2G) facilities; ii) intruders rely on complete network information of IEGSs for designing FDIAs. Table I shows research gaps in [8]-[15] and the explanation for the CAI.

Particularly, research gap ii) poses a significant challenge for designing FDIAs on IEGSs, as intruders may *lack full knowledge of network information*. Previous works [5]-[7] (for power systems), [8]-[15] (for IEGSs), however, assume that intruders have complete network topology and parameters information. In response, researchers investigated cyberattacks on power systems (rather than IEGSs) with incomplete network information [16]-[20]. See [21] for a review.

TABLE I
COMPARISON WITH EXISTING WORKS REGARDING CYBERATTACKS ON IEGSS

| Ref. | Type of attack | Attack target | CAI[#a] | Network information[#b] |
|---|---|---|---|---|
| [8] | FDIA | Gas systems in IEGSs | × | C |
| [9] | FDIA | Gas systems in IEGSs | × | C |
| [10] | Cyberattack | Gas systems in IEGSs | × | C |
| [11] | FDIA | Separate power and gas systems in an IEGS | × | C |
| [12] | LR attack | Power systems in IEGSs | × | C |
| [13] | LR attack | Entire IEGSs[#c] | × | C |
| [14] | FDIA | Entire IEGSs[#c] | × | C |
| [15] | FDIA | Entire integrated electricity-gas-water systems[#c] | × | C |
| This work | FDIA | Entire IEGSs considering power-gas interdependency | ✓ | C and IC |

#a: Cyberattack interdependency (CAI), i.e., if a cyberattack considers coupling constraints, e.g., the gas supply to gas-fired power generators, power-to-gas (P2G) facilities, and/or detailed compressor models.
#b: Intruders are assumed to have complete/incomplete (C/IC) network information.
#c: These works do not consider the interdependency between power and gas systems within an IEGS, although the FDIA target is the entire IEGS/integrated electricity-gas-water systems.

To the best of our knowledge, FDIAs on IEGSs with *incomplete network information* have never been studied, despite the fact that intruders are more likely to have only local network information of an IEGS in the real world. Existing FDIA paradigms on power systems with complete/incomplete network information [5]-[7]/[16]-[20] cannot be applied directly to compromise IEGSs or the power subsystem in an IEGS due to i) significant differences between power and gas infrastructures (i.e., distinct mathematical models) and ii) the CAI of an IEGS.

Considering the importance of large-scale transmission-level gas systems, e.g., European gas transmission networks [22], and IEGSs, this work addresses these critical research gaps and studies FDIAs on *transmission-level* IEGSs. Table I summarizes the major differences between this work and previous works. In essence, the contributions of this work are twofold:

*1) FDIAs on IEGSs with complete network information considering CAI and detailed compressor models*. We develop a tailored static IEGS SE model (based on [23]) and a BDD method to detect bad data during the SE. Then, we propose FDIAs on IEGSs when intruders have complete network information of IEGSs and give conditions to ensure the stealthiness of FDIAs. Particularly, for the first time, the proposed FDIAs consider the critical CAI induced by coupling components, i.e., gas-fired power generators and P2G facilities, significantly distinguishing them from the FDIAs proposed in [8]- [15] that overlook the CAI. We prove that the proposed FDIAs are effective for both simplified and detailed compressor models.

*2) FDIAs on IEGSs with incomplete network information*. For the first time, we develop FDIAs on IEGSs when intruders have only *local network (topology and parameter) information* of IEGSs and give conditions to ensure the stealthiness of these FDIAs. Furthermore, for the first time, we explore FDIAs on IEGSs when intruders have only *local network topology information* of an IEGS, and theoretically prove that: i) topology-only FDIAs on the power system in an IEGS do not exist in general, and ii) counter-intuitively, topology-only FDIAs on the gas system in an IEGS exist, specifically targeting gas compressors. We also reveal two types of FDIA-vulnerable IEGSs, i.e., the IEGS that has nonzero gas load nodes connected by gas compressors, and the IEGS that has meshed gas compressors. Experimental results validate the effectiveness of the proposed FDIAs on IEGSs.

It is expected that this work contributes to the expansion of cyberattack paradigms on the energy sector and aids IEGS operators in better understanding intruders' behavior. The implications of this research are also to alert policymakers and practitioners in the energy sector that IEGSs are vulnerable to FDIAs, so proper detection and mitigation strategies have to be in place to mitigate the adverse effects of FDIAs. The rest of this paper is organized as follows. Section II introduces IEGS SE. Section III designs FDIAs on IEGSs with complete network information. Section IV studies FDIAs on IEGSs with incomplete network information. Section V presents testing results. Section VI draws conclusions.

## II. STATE ESTIMATION AND BAD DATA DETECTION

This section develops a tailored static IEGS SE model (based on [23]) and a BDD method for identifying bad data during the SE. Before proceeding, we clarify the scope of this work.

1) This work aims at connected transmission-level IEGSs, as gas systems are more likely to couple with power systems at a transmission level [22], [24]. We adopt the Weymouth equation [8], [11] to model the gas flow in gas passive pipelines.

2) This work aims at steady-state IEGSs [8]-[11]. The dynamics of IEGSs, e.g., the transit states induced by FDIAs and multiple time scale issues, are another important research topic but are beyond the scope of this paper.

3) This work aims at observable IEGSs (from IEGS operators' perspective). Please refer to Section II.C for the discussion about the observability of IEGSs.

***Notations**:* This paper utilizes i) regular lowercase or uppercase letters to denote scalars; ii) bold lowercase and uppercase letters to denote vectors and matrices, respectively; iii) calligraphic/hollow uppercase letters to denote sets. Vectors are column vectors, whose transpose is denoted by superscript $^{\mathrm{T}}$. The term "FDIAs on power systems/gas systems/IEGSs" denotes "FDIAs targeting the SE of power systems/gas systems/IEGSs". Network information refers to network topology and parameter information. A (gas) compressor refers to the gas pipeline that is driven by a compressor [25].

### A. Power System SE, BDD, and FDIAs: A Preliminary

Power system measurements $\mathbf{z}^{\mathrm{p}}$ and state variables $\mathbf{x}^{\mathrm{p}}$ satisfy

$$\mathbf{z}^{\mathrm{p}} = \mathbf{h}^{\mathrm{p}}(\mathbf{x}^{\mathrm{p}}) + \mathbf{e}^{\mathrm{p}}. \tag{1}$$

$\mathbf{z}^{\mathrm{p}} = col(p_i, q_i, p_{ij}, q_{ij}, v_i, \theta_{i'})$, $i \in \mathcal{P}_n$, $\{i, j\} \in \mathcal{P}_l$, $i' \in \mathcal{P}_n'$, and $\mathbf{x}^{\mathrm{p}} = col(v_i, \theta_i)$, $i \in \mathcal{P}_n$. $col(\cdot)$ maps vectors and/or scalars into one vector, e.g., $col(\mathbf{a}, \mathrm{b}, \mathbf{c}) = [\mathbf{a}^{\mathrm{T}}, \mathrm{b}, \mathbf{c}^{\mathrm{T}}]^{\mathrm{T}}$. $\mathbf{h}^{\mathrm{p}}(\cdot)$ is the functions that map power system state variables into measurements, i.e., (A.1)-(A.4) in Appendix A and identity mappings of $v_i$ and $\theta_{i'}$, $i \in \mathcal{P}_n$, $i' \in \mathcal{P}_n'$. Measurement errors $\mathbf{e}^{\mathrm{p}}$ are assumed to follow $\mathcal{N}(0, \mathbf{E}^{\mathrm{p}})$. By using measurements $\mathbf{z}^{\mathrm{p}}$, system operators can estimate power system states (i.e., *power system SE* [6]) via

$$\hat{\mathbf{x}}^{\mathrm{p}} = \arg\min_{\mathbf{x}^{\mathrm{p}}} (\mathbf{z}^{\mathrm{p}} - \mathbf{h}^{\mathrm{p}}(\mathbf{x}^{\mathrm{p}}))^{\mathrm{T}} \cdot (\mathbf{E}^{\mathrm{p}})^{-1} \cdot (\mathbf{z}^{\mathrm{p}} - \mathbf{h}^{\mathrm{p}}(\mathbf{x}^{\mathrm{p}})), \tag{2}$$

where $\hat{\mathbf{x}}^{\mathrm{p}}$ is the estimated power system states. In previous works [5]-[20], power systems are assumed to be observable.

However, power system measurements may contain bad data due to communication noises and malicious data, affecting the SE [6]. BDD can detect bad data by checking the residuals $\mathbf{r}^{\mathrm{p}}$,

$$\|\mathbf{r}^{\mathrm{p}}\| = \|\mathbf{z}^{\mathrm{p}} - \mathbf{h}^{\mathrm{p}}(\hat{\mathbf{x}}^{\mathrm{p}})\| \le \tau^{\mathrm{p}}. \tag{3}$$

If $\|\mathbf{r}^{\mathrm{p}}\| > \tau^{\mathrm{p}}$, there exist bad data in $\mathbf{z}^{\mathrm{p}}$, where $\tau^{\mathrm{p}}$ is a predefined threshold and $\|\cdot\|$ denotes the $l_2$-norm [6].

Previous works [6], [18] show that under some assumptions, FDIAs can bypass power system BDD by injecting sophisticated false data into measurements. Specifically, let $\Delta\mathbf{z}^{\mathrm{p}} = col(\Delta p_i, \Delta q_i, \Delta p_{ij}, \Delta q_{ij}, \Delta v_i, \Delta\theta_{i'})$, $i \in \mathcal{P}_n$, $\{i,j\} \in \mathcal{P}_l$, $i' \in \mathcal{P}_n'$, and $\Delta\mathbf{x}^{\mathrm{p}} = col(\Delta v_i, \Delta\theta_i)$, $i \in \mathcal{P}_n$ be the injected false data and the variations in estimated states after FDIAs, respectively. If

$$\Delta\mathbf{z}^{\mathrm{p}} = \mathbf{h}^{\mathrm{p}}(\hat{\mathbf{x}}^{\mathrm{p}} + \Delta\mathbf{x}^{\mathrm{p}}) - \mathbf{h}^{\mathrm{p}}(\hat{\mathbf{x}}^{\mathrm{p}}) \tag{4}$$

holds, the FDIA $\Delta\mathbf{z}^{\mathrm{p}}$ designed based on (4) can bypass BDD and compromise power systems *stealthily*.

### *B. Gas System SE: A New Definition of State Variables*

This subsection develops gas system SE. First, we give

***Definition 1*** *(Measurements and State Variables in Gas Systems):* This paper defines $\mathbf{z}^{\mathrm{g}}$ and $\mathbf{x}^{\mathrm{g}}$ as the measurements and state variables in a gas system, respectively, where $\mathbf{z}^{\mathrm{g}} = col(g_i, g_{ij}, c_{mn}, \pi_i)$, $i \in \mathcal{G}_n$, $\{i,j\} \in \mathcal{G}_l$, $\{m, n\} \in \mathcal{G}_c$, and $\mathbf{x}^{\mathrm{g}} = col(c_{ij}, \pi_i)$, $\{i,j\} \in \mathcal{G}_c$, $i \in \mathcal{G}_n$.

Gas system measurements and state variables satisfy

$$\mathbf{z}^{\mathrm{g}} = \mathbf{h}^{\mathrm{g}}(\mathbf{x}^{\mathrm{g}}) + \mathbf{e}^{\mathrm{g}}, \tag{5}$$

where $\mathbf{h}^{\mathrm{g}}(\cdot)$ is the functions that map gas system state variables into measurements, i.e., (A.5)-(A.6) in Appendix A and identity mappings of $c_{ij}$ and $\pi_i$, $\{i,j\} \in \mathcal{G}_c$, $i \in \mathcal{G}_n$. $\mathbf{e}^{\mathrm{g}} \sim \mathcal{N}(0, \mathbf{E}^{\mathrm{g}})$ is the measurement errors. With measurements $\mathbf{z}^{\mathrm{g}}$, gas system states are estimated by

$$\hat{\mathbf{x}}^{\mathrm{g}} = \arg\min_{\mathbf{x}^{\mathrm{g}}} (\mathbf{z}^{\mathrm{g}} - \mathbf{h}^{\mathrm{g}}(\mathbf{x}^{\mathrm{g}}))^{\mathrm{T}} \cdot (\mathbf{E}^{\mathrm{g}})^{-1} \cdot (\mathbf{z}^{\mathrm{g}} - \mathbf{h}^{\mathrm{g}}(\mathbf{x}^{\mathrm{g}})), \tag{6}$$

where $\hat{\mathbf{x}}^{\mathrm{g}}$ is the estimated gas system states. Previous works [8]-[11], [26] assume that gas systems are observable, which is also adopted in this work.

***Remark 1:*** Unlike [23], [26], this work adopts Definition 1 to define gas system state variables $\mathbf{x}^{\mathrm{g}}$ so that they can *uniquely* determine the values of all gas system variables, avoiding inconsistent estimation of measurements in the gas system. Hereinafter, we adopt Definition 1 and gas system SE (6) as the basis for developing IEGS SE and studying FDIAs on IEGEs.

### *C. Tailored IEGS SE and BDD*

Based on [23], [27], and Definition 1, we develop a tailored static IEGS SE. A significant distinction between the SE of IEGSs and pure power/gas systems lies in the consideration of power-gas interdependency (i.e., gas-fired power generators and P2G facilities). The next section shows how the interdependency affects FDIAs on IEGSs. First, we give

***Definition 2*** *(Measurements and State Variables in IEGSs):* This paper defines $\mathbf{z} = col(\mathbf{z}^{\mathrm{p}}, \mathbf{z}^{\mathrm{g}})$ and $\mathbf{x} = col(\mathbf{x}^{\mathrm{p}}, \mathbf{x}^{\mathrm{g}})$ as the measurements and state variables in an IEGS, respectively.

Measurements and state variables in an IEGS satisfy

$$\mathbf{z} = \mathbf{h}(\mathbf{x}) + \mathbf{e}, \tag{7}$$

where $\mathbf{h}(\cdot) = col(\mathbf{h}^{\mathrm{p}}(\cdot), \mathbf{h}^{\mathrm{g}}(\cdot))$, and $\mathbf{e} \sim \mathcal{N}(0, \mathbf{E})$ is the measurement errors. In view of the interdependency between power and gas subsystems in an IEGS (see (A.7)-(A.11) in Appendix A for details), the estimated IEGS states $\hat{\mathbf{x}}$ are derived by

$$\hat{\mathbf{x}} = \arg\min_{\mathbf{x}} (\mathbf{z} - \mathbf{h}(\mathbf{x}))^{\mathrm{T}} \cdot \mathbf{E}^{-1} \cdot (\mathbf{z} - \mathbf{h}(\mathbf{x})) \tag{8a}$$

$$\text{s.t. } \mathbf{T} \cdot \mathbf{h}_{\mathrm{c}}^{\mathrm{p}}(\mathbf{x}^{\mathrm{p}}) = \mathbf{h}_{\mathrm{c}}^{\mathrm{g}}(\mathbf{x}^{\mathrm{g}}), \tag{8b}$$

where constraint (8b) is the vector form of (A.10) and (A.11). $\mathbf{T}$ is a diagonal matrix, whose diagonal elements are $\gamma_g$, $g \in \mathcal{G}_g$, and $\gamma_f^{-1}$, $f \in \mathcal{P}_f$.

***Remark 2:*** Physically, the outputs of gas-fired power generators/P2G facilities that satisfy assumptions A.i)-A.iii)/B.i)-B.iii) (in Appendix A) can be estimated by either power system state variables $\mathbf{x}^{\mathrm{p}}$, i.e., $\mathbf{h}_{\mathrm{c}}^{\mathrm{p}}(\mathbf{x}^{\mathrm{p}})$ via (A.1), (A.3), and (A.7), or gas system state variables $\mathbf{x}^{\mathrm{g}}$, i.e., $\mathbf{T}^{-1} \cdot \mathbf{h}_{\mathrm{c}}^{\mathrm{g}}(\mathbf{x}^{\mathrm{g}})$ via (A.5), (A.6), and (A.9). Constraint (8b) ensures the consistency between the estimated outputs of gas-fired power generators and P2G facilities, i.e., $\mathbf{T} \cdot \mathbf{h}_{\mathrm{c}}^{\mathrm{p}}(\hat{\mathbf{x}}^{\mathrm{p}}) = \mathbf{h}_{\mathrm{c}}^{\mathrm{g}}(\hat{\mathbf{x}}^{\mathrm{g}})$. In other words, without this constraint, the inconsistency cannot be guaranteed and may lead to different estimated values. To this end, constraint (8b) is distinct from the auxiliary constraints aiming at enhancing the accuracy of SE, e.g., zero injection constraints [29], and is indispensable for IEGS SE. Since we assume that IEGSs are observable, model (8) determines all state variables $\mathbf{x}$ (using measurements $\mathbf{z}$). This assumption is based on the fact that both power and gas systems have sufficient meters for measurement collection [5]-[20], [26], yielding observable IEGSs.

In order to detect bad data in IEGS measurements $\mathbf{z}$ during the IEGS SE, we develop the following BDD:

$$\|\mathbf{r}\| = \|\mathbf{z} - \mathbf{h}(\hat{\mathbf{x}})\| \le \tau, \tag{9a}$$

$$\|\mathbf{r}_{\mathrm{c}}\| = \|\mathbf{T} \cdot \mathbf{h}_{\mathrm{c}}^{\mathrm{p}}(\hat{\mathbf{x}}^{\mathrm{p}}) - \mathbf{h}_{\mathrm{c}}^{\mathrm{g}}(\hat{\mathbf{x}}^{\mathrm{g}})\| \le \epsilon. \tag{9b}$$

Vector $\mathbf{r}$ and $\mathbf{r}_{\mathrm{c}}$ are residuals for all IEGS and coupling measurements, respectively. If $\|\mathbf{r}\|$ exceeds a threshold $\tau$ or $\|\mathbf{r}_{\mathrm{c}}\|$ exceeds a small positive number $\epsilon$, e.g., 1e-5 (for numerical consideration), bad data exist in measurements $\mathbf{z}$. Constraint (9a) is similar to the BDD in power systems, while constraint (9b) is tailored for IEGS BDD to avoid inconsistent estimation and detect the FDIAs that ignore coupling constraint (8b).

## III. FDIAs on IEGSs With Complete Network Information

In this section, we propose FDIAs on IEGSs with complete network information.

### *A. The Proposed FDIAs on IEGSs*

By referring to FDIAs on power systems [5], we give the definition of FDIAs on IEGSs.

***Definition 3*** *(FDIAs on IEGSs):* If intruders can attack and compromise IEGS SE by injecting malicious data into IEGS measurements without being detected by IEGS BDD (9), these attacks are defined as FDIAs on IEGSs.

Based on Definition 3, this subsection proposes FDIAs on IEGSs. Note that existing works in terms of cyberattacks on IEGSs [8]-[15] do not consider CAI caused by coupling constraints (8b) and are detectable by IEGS BDD (9). Differently, the proposed FDIAs consider the CAI and thus can bypass (9). In this subsection, we assume that i) true measurements $\mathbf{z}$ can pass IEGS BDD, ii) intruders have full knowledge of network information $\mathbf{h}(\cdot)$, iii) intruders can get access to measurements $\mathbf{z}$, and iv) intruders have the same estimated states $\hat{\mathbf{x}}$ as those estimated by IEGS operators. The above assumptions are also adopted in [8]-[15].

For the proposed FDIAs on IEGSs, let $\Delta\mathbf{z} = col(\Delta\mathbf{z}^{\mathrm{p}}, \Delta\mathbf{z}^{\mathrm{g}})$ and $\Delta\mathbf{x} = col(\Delta\mathbf{x}^{\mathrm{p}}, \Delta\mathbf{x}^{\mathrm{g}})$ be the injected false data into measurements $\mathbf{z}$ and the variations in estimated states $\hat{\mathbf{x}}$ after the attack, respectively. $\Delta\mathbf{z}^{\mathrm{g}} = col(\Delta g_i, \Delta g_{ij}, \Delta c_{mn}, \Delta\pi_i)$, $i \in \mathcal{G}_n$, $\{i, j\} \in \mathcal{G}_l$, $\{m, n\} \in \mathcal{G}_c$, is the injected false data into measurements $\mathbf{z}^{\mathrm{g}}$. $\Delta\mathbf{x}^{\mathrm{g}} = col(\Delta c_{ij}, \Delta\pi_i)$, $\{i, j\} \in \mathcal{G}_c$, $i \in \mathcal{G}_n$, is the variations in estimated states $\hat{\mathbf{x}}^{\mathrm{g}}$ after the attack. If an attack $\Delta\mathbf{z}$ satisfies

$$\Delta\mathbf{z} = \mathbf{h}(\hat{\mathbf{x}} + \Delta\mathbf{x}) - \mathbf{h}(\hat{\mathbf{x}}), \tag{10a}$$

$$\|\mathbf{T} \cdot \mathbf{h}_{\mathrm{c}}^{\mathrm{p}}(\hat{\mathbf{x}}^{\mathrm{p}} + \Delta\mathbf{x}^{\mathrm{p}}) - \mathbf{h}_{\mathrm{c}}^{\mathrm{g}}(\hat{\mathbf{x}}^{\mathrm{g}} + \Delta\mathbf{x}^{\mathrm{g}})\| \le \epsilon, \tag{10b}$$

let $\mathbf{z}_{\mathrm{bad}} = \mathbf{z} + \Delta\mathbf{z}$ and $\mathbf{x}_{\mathrm{bad}} = \hat{\mathbf{x}} + \Delta\mathbf{x}$, where $\mathbf{z}_{\mathrm{bad}}$ and $\mathbf{x}_{\mathrm{bad}}$ are falsified measurements and estimated states under falsified measurements $\mathbf{z}_{\mathrm{bad}}$, respectively. We have

$$\begin{aligned} \|\mathbf{r}_{\mathrm{bad}}\| &= \|\mathbf{z}_{\mathrm{bad}} - \mathbf{h}(\mathbf{x}_{\mathrm{bad}})\| \\ &= \|\mathbf{z} + \Delta\mathbf{z} - \mathbf{h}(\hat{\mathbf{x}} + \Delta\mathbf{x})\| \\ &= \|\mathbf{z} + \mathbf{h}(\hat{\mathbf{x}} + \Delta\mathbf{x}) - \mathbf{h}(\hat{\mathbf{x}}) - \mathbf{h}(\hat{\mathbf{x}} + \Delta\mathbf{x})\| \\ &= \|\mathbf{z} - \mathbf{h}(\hat{\mathbf{x}})\| = \|\mathbf{r}\| \le \tau, \end{aligned} \tag{11a}$$

$$\begin{aligned} \|\mathbf{r}_{\mathrm{c,\,bad}}\| &= \|\mathbf{T} \cdot \mathbf{h}_{\mathrm{c}}^{\mathrm{p}}(\mathbf{x}_{\mathrm{bad}}^{\mathrm{p}}) - \mathbf{h}_{\mathrm{c}}^{\mathrm{g}}(\mathbf{x}_{\mathrm{bad}}^{\mathrm{g}})\| \\ &= \|\mathbf{T} \cdot \mathbf{h}_{\mathrm{c}}^{\mathrm{p}}(\hat{\mathbf{x}}^{\mathrm{p}} + \Delta\mathbf{x}^{\mathrm{p}}) - \mathbf{h}_{\mathrm{c}}^{\mathrm{g}}(\hat{\mathbf{x}}^{\mathrm{g}} + \Delta\mathbf{x}^{\mathrm{g}})\| \le \epsilon, \end{aligned} \tag{11b}$$

where $\mathbf{r}_{\mathrm{bad}}$ and $\mathbf{r}_{\mathrm{c,\,bad}}$ are the residuals for all falsified IEGS measurements and coupling measurements, respectively. $\mathbf{x}_{\mathrm{bad}}^{\mathrm{g}} = \hat{\mathbf{x}}^{\mathrm{g}} + \Delta\mathbf{x}^{\mathrm{g}}$. According to (11), the attack $\Delta\mathbf{z}$ can compromise an IEGS without being detected by IEGS BDD (9) and is an FDIA.

***Remark 3**:* Conditions (10) ensure that the residuals of IEGS BDD (after FDIAs) are within thresholds $\tau$ and $\epsilon$, respectively. Particularly, condition (10b) is the tailored product of the CAI (induced by coupling constraint (8b)) and is critical for FDIAs to bypass IEGS BDD (9b), distinguishing the proposed FDIAs from existing FDIAs [5]-[15]. In Section V.A, we further exhibit the value of (10b) by a numerical test, where the FDIAs on pure power systems cannot be applied to compromise even the power subsystem in an IEGS.

Based on the above analysis, we immediately derive

***Proposition 1**:* Given assumptions i)-iv) (the 3rd paragraph in this subsection), intruders can design FDIAs $\Delta\mathbf{z}$ on SE (8) of transmission-level IEGSs with complete network (topology and parameter) information (i.e., $\mathbf{h}(\cdot)$) by following conditions (10).

***Remark 4**:* Proposition 1 is derived based on assumptions i)-iv) presented in this subsection. The first assumption is easy to be satisfied. For assumptions ii) and iii), they are widely adopted in previous works [8]-[15]. In the real world, although strict, these two assumptions can still be satisfied during insider attacks, e.g., the cyberattack on Australian wastewater services company [30]. Another significance of assumptions ii) and iii) is to help IEGS operators evaluate the vulnerability of IEGSs to FDIAs [5]. In the next section, we generalize Proposition 1 and study the cases when assumptions ii) and iii) do not hold. For the last assumption, which is broadly adopted [8]-[15], it may not always be true [31]. The above analysis is based on the assumption that intruders derive the same estimated states as those estimated by IEGS operators. In practice, this may not always be true [31]. We analyze the case when this condition does not hold.

Without loss of generality, let $\hat{\mathbf{x}}' = \hat{\mathbf{x}} + \boldsymbol{\xi}$, where $\boldsymbol{\xi}$ is the *bias* between $\hat{\mathbf{x}}'$ (estimated states by intruders) and $\hat{\mathbf{x}}$ (estimated states by IEGS operators). Next, we analyze the impact of bias on FDIAs on IEGSs by observing residuals $\mathbf{r}_{\mathrm{bad}}$ and $\mathbf{r}_{\mathrm{c,\,bad}}$. Since intruders do not know the existence or the exact value of the bias in their estimated states, they are expected to routinely employ conditions (10) to design an FDIA, i.e.,

$$\Delta\mathbf{z} = \mathbf{h}(\hat{\mathbf{x}} + \boldsymbol{\xi} + \Delta\mathbf{x}) - \mathbf{h}(\hat{\mathbf{x}} + \boldsymbol{\xi}), \tag{12a}$$

$$\|\mathbf{T} \cdot \mathbf{h}_{\mathrm{c}}^{\mathrm{p}}(\hat{\mathbf{x}}^{\mathrm{p}} + \boldsymbol{\xi}^{\mathrm{p}} + \Delta\mathbf{x}^{\mathrm{p}}) - \mathbf{h}_{\mathrm{c}}^{\mathrm{g}}(\hat{\mathbf{x}}^{\mathrm{g}} + \boldsymbol{\xi}^{\mathrm{g}} + \Delta\mathbf{x}^{\mathrm{g}})\| \le \epsilon_{\mathrm{bad}}, \tag{12b}$$

where $\boldsymbol{\xi} = col(\boldsymbol{\xi}^{\mathrm{p}}, \boldsymbol{\xi}^{\mathrm{g}})$. Then, IEGS operators conduct BDD (11) by examining $\|\mathbf{r}_{\mathrm{bad}}\|$ and $\|\mathbf{r}_{\mathrm{c,\,bad}}\|$. By replacing $\Delta\mathbf{z}$ in (11a) with (12a), we have

$$\begin{aligned} \|\mathbf{r}_{\mathrm{bad}}\| &= \|\mathbf{z} - \mathbf{h}(\hat{\mathbf{x}}) + \mathbf{h}(\hat{\mathbf{x}} + \boldsymbol{\xi} + \Delta\mathbf{x}) - \mathbf{h}(\hat{\mathbf{x}} + \boldsymbol{\xi}) + \mathbf{h}(\hat{\mathbf{x}}) - \mathbf{h}(\hat{\mathbf{x}} + \Delta\mathbf{x})\| \\ &= \|\mathbf{r} + (\mathbf{h}(\hat{\mathbf{x}} + \Delta\mathbf{x} + \boldsymbol{\xi}) - \mathbf{h}(\hat{\mathbf{x}} + \Delta\mathbf{x})) - (\mathbf{h}(\hat{\mathbf{x}} + \boldsymbol{\xi}) - \mathbf{h}(\hat{\mathbf{x}}))\|. \end{aligned} \tag{13}$$

Equation (13) quantitatively analyzes the impact of bias $\boldsymbol{\xi}$ on $\|\mathbf{r}_{\mathrm{bad}}\|$. This enables intruders to *exactly* evaluate the stealthiness of any FDIA under different biases. Particularly, when $\boldsymbol{\xi}$ is small, i.e., intruders have confidence in their estimated states, we have

$$\begin{aligned} \|\mathbf{r}_{\mathrm{bad}}\| &= \|\mathbf{r} + (\mathbf{h}(\hat{\mathbf{x}} + \Delta\mathbf{x} + \boldsymbol{\xi}) - \mathbf{h}(\hat{\mathbf{x}} + \Delta\mathbf{x})) - (\mathbf{h}(\hat{\mathbf{x}} + \boldsymbol{\xi}) - \mathbf{h}(\hat{\mathbf{x}}))\| \\ &= \|\mathbf{r} + (\mathbf{J}_1 - \mathbf{J}_2) \cdot \boldsymbol{\xi} + o(\boldsymbol{\xi})\| \\ &\approx \|\mathbf{r} + (\mathbf{J}_1 - \mathbf{J}_2) \cdot \boldsymbol{\xi}\| \\ &\le \|\mathbf{r}\| + \|(\mathbf{J}_1 - \mathbf{J}_2)\| \cdot \|\boldsymbol{\xi}\|, \end{aligned} \tag{14}$$

an *approximate upper bound* for the FDIAs on IEGSs. $\mathbf{J}_1 = \partial\mathbf{h}/\partial\mathbf{x}|_{\mathbf{x} = \hat{\mathbf{x}} + \Delta\mathbf{x}}$ and $\mathbf{J}_2 = \partial\mathbf{h}/\partial\mathbf{x}|_{\mathbf{x} = \hat{\mathbf{x}}}$ are constant matrices. The second line in (14) is derived by the Taylor expansion, and the third line is an approximation by neglecting higher-order terms of $\boldsymbol{\xi}$, as $\boldsymbol{\xi}$ is assumed to be small. If this bound is no larger than $\tau$, we have $\|\mathbf{r}_{\mathrm{bad}}\| \le \tau$, i.e., $\|\mathbf{r}_{\mathrm{bad}}\|$ can bypass BDD. Note that the last line in (14) indicates that the upper bound of $\|\mathbf{r}_{\mathrm{bad}}\|$ varies affinely with bias $\boldsymbol{\xi}$. This affine relation allows intruders to easily extrapolate how the variation of a bias affects $\|\mathbf{r}_{\mathrm{bad}}\|$, so as to evaluate the stealthiness of an FDIA.

Different from $\|\mathbf{r}_{\mathrm{bad}}\|$, the BDD for $\|\mathbf{r}_{\mathrm{c,\,bad}}\|$ only allows a very small numerical error (e.g., $\epsilon_{\mathrm{bad}} = 1e^{-5}$). Any bias in estimated states $\hat{\mathbf{x}}'$ may result in the violation of (11b), as the FDIA is designed based on (12b). In order to enable $\|\mathbf{r}_{\mathrm{c,\,bad}}\|$ to bypass BDD, intruders should manage to ensure the accurate estimation for the state variables $\mathbf{x}$ in (A.10) and (A.11)[1], i.e.,

$$\mathbf{A}^{\mathrm{p}} \cdot \boldsymbol{\xi}^{\mathrm{p}} = 0,\ \mathbf{A}^{\mathrm{g}} \cdot \boldsymbol{\xi}^{\mathrm{g}} = 0. \tag{15}$$

$\mathbf{A}^{\mathrm{p}}$ and $\mathbf{A}^{\mathrm{g}}$ are coefficient matrices, where $\mathbf{A}^{\mathrm{p}} \cdot \boldsymbol{\xi}^{\mathrm{p}}$ and $\mathbf{A}^{\mathrm{g}} \cdot \boldsymbol{\xi}^{\mathrm{g}}$ denote the errors of the estimated power and gas system states in (A.10) and (A.11), respectively. For this case, condition (12b) degenerates to condition (10b), i.e.,

$$\|\mathbf{r}_{\mathrm{c,\,bad}}\| = \|\mathbf{T} \cdot \mathbf{h}_{\mathrm{c}}^{\mathrm{p}}(\hat{\mathbf{x}}^{\mathrm{p}} + \Delta\mathbf{x}^{\mathrm{p}}) - \mathbf{h}_{\mathrm{c}}^{\mathrm{g}}(\hat{\mathbf{x}}^{\mathrm{g}} + \Delta\mathbf{x}^{\mathrm{g}})\|$$

[1] The state variables in in (A.10) and (A.11) refer to the state variables $\mathbf{x}$ that are related to the power and gas injections in (A.10) and (A.11).

$$= \|\mathbf{T}\cdot\mathbf{h}_{\mathrm{c}}^{\mathrm{p}}(\hat{\mathbf{x}}^{\mathrm{p}} + \boldsymbol{\xi}^{\mathrm{p}} + \mathbf{x}^{\mathrm{p}}) - \mathbf{h}_{\mathrm{c}}^{\mathrm{g}}(\hat{\mathbf{x}}^{\mathrm{g}} + \boldsymbol{\xi}^{\mathrm{g}} + \Delta\mathbf{x}^{\mathrm{g}})\| \leq \epsilon_{\mathrm{bad}},$$

indicating $\|\mathbf{r}_{\mathrm{c,\,bad}}\|$ can bypass BDD. In practice, condition (15) is elusive, as intruders hardly know exact biases. Another method is to keep the estimation of the state variables in (A.10) and (A.11) unchanged (before and after an FDIA), i.e.,

$$\mathbf{A}^{\mathrm{p}}\cdot\Delta\mathbf{x}^{\mathrm{p}} = 0,\ \mathbf{A}^{\mathrm{g}}\cdot\Delta\mathbf{x}^{\mathrm{g}} = 0. \quad (16)$$

In this case, intruders employ (16) instead of (12b) for keeping relevant estimated states the same. Thus, $\|\mathbf{r}_{\mathrm{c,\,bad}}\| = \|\mathbf{T}\cdot\mathbf{h}_{\mathrm{c}}^{\mathrm{p}}(\hat{\mathbf{x}}^{\mathrm{p}} + \Delta\mathbf{x}^{\mathrm{p}}) - \mathbf{h}_{\mathrm{c}}^{\mathrm{g}}(\hat{\mathbf{x}}^{\mathrm{g}} + \Delta\mathbf{x}^{\mathrm{g}})\| = \|\mathbf{T}\cdot\mathbf{h}_{\mathrm{c}}^{\mathrm{p}}(\hat{\mathbf{x}}^{\mathrm{p}}) - \mathbf{h}_{\mathrm{c}}^{\mathrm{g}}(\hat{\mathbf{x}}^{\mathrm{g}})\| = 0 \leq \epsilon_{\mathrm{bad}}$, i.e., the $\|\mathbf{r}_{\mathrm{c,\,bad}}\|$ can bypass BDD.

### *B. Stealthiness Enhancement*

The stealthiness of the proposed FDIAs on IEGSs (in Section III.A) can be further extended in the following aspects.

*1) Operation constraints.* Although conditions (10) ensure the stealthiness of FDIAs on IEGSs for bypassing (general) IEGS BDD (9), these FDIAs may still induce the violation of IEGS operation constraints and thus arouse suspicion. In reality, intruders can consider the following operation constraints to enhance the stealthiness of the proposed FDIAs on IEGSs, which enables FDIAs to bypass IEGS BDD and obey IEGS operation constraints simultaneously.

$$\sqrt{(\hat{\mathrm{p}}_{ij} + \Delta p_{ij})^2 + (\hat{\mathrm{q}}_{ij} + \Delta q_{ij})^2} \leq \mathrm{S}_{ij}^{\max},\ \forall\{i, j\} \in \mathcal{P}_l, \quad (17a)$$

$$\mathrm{V}_i^{\min} \leq \hat{\mathrm{v}}_i + \Delta v_i \leq \mathrm{V}_i^{\max},\ \ \forall i \in \mathcal{P}_n, \quad (17b)$$

$$-\theta_i^{\max} \leq \hat{\theta}_i + \Delta\theta_i \leq \theta_i^{\max},\ \ \forall i \in \mathcal{P}_n, \quad (17c)$$

$$\mathrm{P}_i^{\min} \leq \hat{\mathrm{p}}_i + \Delta p_i \leq \mathrm{P}_i^{\max},\ \forall i \in \mathcal{P}_n, \quad (17d)$$

$$\mathrm{Q}_i^{\min} \leq \hat{\mathrm{q}}_i + \Delta q_i \leq \mathrm{Q}_i^{\max},\ \forall i \in \mathcal{P}_n, \quad (17e)$$

$$-\mathrm{G}_{ij}^{\max} \leq \hat{\mathrm{g}}_{ij} + \Delta g_{ij} \leq \mathrm{G}_{ij}^{\max},\ \ \forall\{i, j\} \in \mathcal{G}_l, \quad (17f)$$

$$0 \leq \hat{\mathrm{c}}_{ij} + \Delta c_{ij} \leq \mathrm{C}_{ij}^{\max},\ \ \forall\{i, j\} \in \mathcal{G}_c, \quad (17g)$$

$$\Pi_i^{\min} \leq \hat{\pi}_i + \Delta\pi_i \leq \Pi_i^{\max},\ \ \forall i \in \mathcal{G}_n, \quad (17h)$$

$$\hat{\pi}_j + \Delta\pi_j \leq \alpha_{ij}(\hat{\pi}_i + \Delta\pi_i),\ \ \forall\{i, j\} \in \mathcal{G}_c, \quad (17i)$$

$$\mathrm{G}_i^{\min} \leq \hat{\mathrm{g}}_i + \Delta g_i \leq \mathrm{G}_i^{\max},\ \forall i \in \mathcal{G}_n. \quad (17j)$$

$\hat{\mathrm{p}}_{ij}$, $\hat{\mathrm{q}}_{ij}$, $\hat{\mathrm{g}}_{ij}$, $\hat{\mathrm{p}}_i$, $\hat{\mathrm{q}}_i$, $\hat{\mathrm{g}}_i$ are the elements in estimated measurements $\mathbf{h}(\hat{\mathbf{x}})$, and $\hat{\mathrm{v}}_i$, $\hat{\theta}_i$, $\hat{\mathrm{c}}_{ij}$, $\hat{\pi}_i$ are the elements in estimated states $\hat{\mathbf{x}}$. Constraints (17a)-(17j) enforce power flows, voltage magnitudes, phase angles, real and reactive power injections, gas flow in gas passive pipelines and gas compressors, pressures, gas compressors, and gas injections, respectively.

*2) Distance-based BDD.* Although the proposed FDIAs can bypass IEGS BDD (9), they may still be detected by distance-based BDD, such as machine learning methods, as discussed in [32]. The reason is that false data can be classified as "outliers" due to the greater distance from normal data. In order to enhance the stealthiness of the proposed FDIAs for bypassing distance-based BDD, we adopt the mechanism proposed in [32], i.e., reducing the distance of false data from normal measurements by adding the following objective function to constraints (10).

$$\min_{\Delta\mathbf{z}} \sum_{i=1}^{N} (\mathbf{z} + \Delta\mathbf{z} - \mathbf{z}_i) \quad (18)$$

$\mathbf{z}_1, \ldots, \mathbf{z}_N$ are historical measurements. Objective function (18) aims to minimize the distance between false data and historical measurements. The effectiveness of this method is validated by [32] for bypassing distance-based BDD.

*3) Detailed compressor model.* The FDIAs proposed in Section III.A are based on a simplified gas compressor model [33]. This model is broadly adopted in large-scale gas transmission networks. For small- and medium-scale IEGSs, e.g., gas distribution network, the simplified compressor model may not be accurate enough to model the energy consumption of a gas compressor and the relation between gas flow and nodal pressure [33]. Detailed compressor models are needed for designing tailored FDIAs (on small- and medium-scale IEGSs). Please find the detailed compressor model in Appendix B.

If we follow Definition 2, accordingly, we define the state variables for turbine and piston compressors as $\mathbf{x}_{\mathrm{T}_ij}^{\mathrm{g}}$ and $\mathbf{x}_{\mathrm{P}_ij}^{\mathrm{g}}$, respectively. For IEGSs with detailed compressor models, we derive an augmented IEGS SE (in Appendix B) by modifying IEGS SE (8). Unfortunately, intruders can still design tailored FDIAs by following an augmented Proposition 1, where all $\mathbf{h}(\cdot)$, $\mathbf{h}_{\mathrm{c}}^{\mathrm{p}}(\cdot)$, $\mathbf{h}_{\mathrm{c}}^{\mathrm{g}}(\cdot)$, $\mathbf{T}$, $\mathbf{x}$, $\mathbf{x}^{\mathrm{g}}$, $\hat{\mathbf{x}}$, $\Delta\mathbf{x}$, and $\Delta\mathbf{x}^{\mathrm{g}}$ in Proposition 1 are replaced with $\underline{\mathbf{h}}(\cdot)$, $\underline{\mathbf{h}}_{\mathrm{c}}^{\mathrm{p}}(\cdot)$, $\underline{\mathbf{h}}_{\mathrm{c}}^{\mathrm{g}}(\cdot)$, $\underline{\mathbf{T}}$, $\underline{\mathbf{x}}$, $\underline{\mathbf{x}}^{\mathrm{g}}$, $\hat{\underline{\mathbf{x}}}$, $\Delta\underline{\mathbf{x}}$, and $\Delta\underline{\mathbf{x}}^{\mathrm{g}}$, respectively. (See Appendix B for their definitions.) Namely, by a simple modification, the FDIAs developed in Section III.A can compromise the IEGSs with detailed compressor models.

## IV. FDIAs on IEGSs With Incomplete Network Information

In practice, intruders may have difficulties in deriving complete network information. For example, intruders in the 2015 Ukraine cyberattack hacked the control center of the power grid in the Ivano-Frankivsk region [34], indicating that the intruders could only get access to the *incomplete network information* of the Ukrainian power grid. For this type of practical situation, this section relaxes the strong assumption adopted in Section III, which requires intruders have complete network (topology and parameter) information, and develops FDIAs on IEGSs with incomplete network information (to be precise, with local network (topology and parameter) information (Section IV.A) and local topology information (Section IV.B)).

### *A. FDIAs on IEGSs With Local Network Information*

Without loss of generality, this subsection divides a connected IEGS into two regions, i.e., an attacking region, where intruders have the network information, and a non-attacking region (the remaining region). Each region may consist of several areas in an IEGS without direct connectivity requirement. The direct connectivity between two areas means that there exists at least one tie line, at least one gas-fired generator, or at least one P2G facility that physically connects these two areas. Fig. 1 shows an example. The attacking/nonattacking region consists of A1/N1 (an area of the power subsystem in an IEGS) and A2/N2 (an area of the gas subsystem in an IEGS). Power transmission line $\{i, j\}$, gas passive pipeline $\{m, n\}$, and gas compressor $\{u, v\}$ are tie-lines that connect the attacking region and non-attacking region. Generally, we have

***Definition 4** (Boundary nodes):* Boundary nodes are defined as those power buses and gas nodes *in attacking regions*: i) the power buses and gas nodes that connect to tie-lines, ii) the power buses with gas-fired generators and the power buses that connect to P2G facilities, and iii) the gas nodes with P2G facilities and the gas nodes that connect to gas-fired generators.

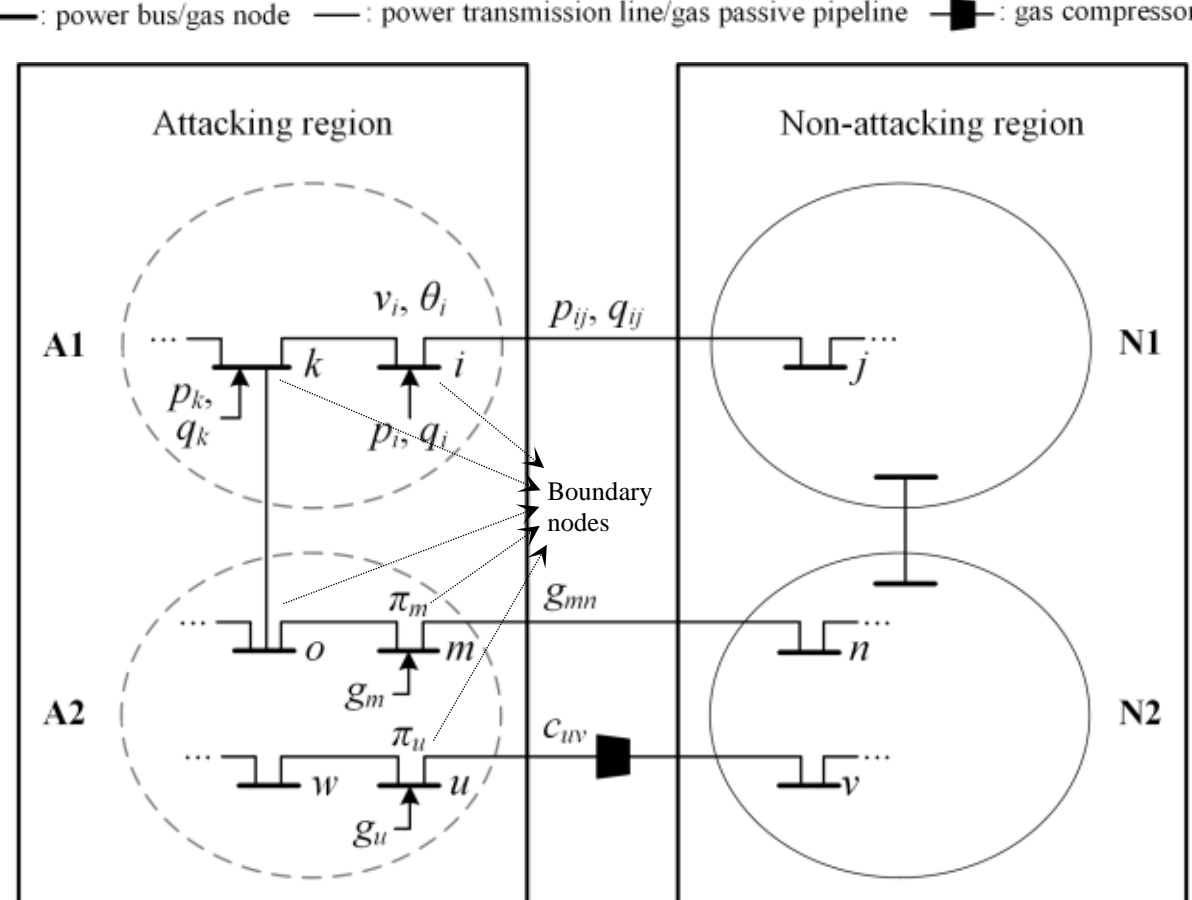

Fig. 1. Illustrative example of attacking and non-attacking regions in an IEGS.

Boundary nodes are critical for developing FDIAs on IEGSs with local network information. Based on Definition 4, in Fig. 1, power bus $i$ and gas nodes $m$ and $u$ are the first type of boundary nodes, and power bus $k$ and gas node $o$ are the second and third types, respectively. Let $\mathbf{z} = col(\mathbf{z}_A, \mathbf{z}_B, \mathbf{z}_N)$ and $\mathbf{x} = col(\mathbf{x}_A, \mathbf{x}_B, \mathbf{x}_N)$. Vector $\mathbf{z}_A$ is composed of all measurements in the attacking region except for power and gas injection measurements at boundary nodes, vector $\mathbf{z}_B$ contains power and gas injection measurements at boundary nodes, and vector $\mathbf{z}_N$ consists of the remaining measurements outside the attacking region. Power and gas flow measurements in tie lines between attacking and non-attacking regions, e.g., $p_{ij}$, $q_{ij}$, $g_{mn}$, and $c_{uv}$ in Fig. 1, belong to $\mathbf{z}_N$. Vector $\mathbf{x}_A$ includes state variables in the attacking region except for state variables at boundary nodes, vector $\mathbf{x}_B$ contains state variables in boundary nodes, and vector $\mathbf{x}_N$ is the remaining state variables outside the attacking region. Gas flow states in gas compressors, e.g., $c_{uv}$ in Fig. 1, belong to $\mathbf{x}_N$. Based on this partition, we (equivalently) rewrite the correlation between estimated measurements and estimated states, i.e., (7), as follows.

$$\begin{bmatrix}\mathbf{z}_A\\ \mathbf{z}_B\\ \mathbf{z}_N\end{bmatrix} = \begin{bmatrix}\mathbf{h}_A(\mathbf{x}_A,\mathbf{x}_B)\\ \mathbf{h}_B^A(\mathbf{x}_A,\mathbf{x}_B)+\mathbf{h}_B^N(\mathbf{x}_N,\mathbf{x}_B)\\ \mathbf{h}_N(\mathbf{x}_N,\mathbf{x}_B)\end{bmatrix} + \begin{bmatrix}\mathbf{e}_A\\ \mathbf{e}_B\\ \mathbf{e}_N\end{bmatrix}. \qquad (19)$$

$\mathbf{h}_A(\cdot)$, $\mathbf{h}_B^A(\cdot)$, $\mathbf{h}_B^N(\cdot)$, $\mathbf{h}_N(\cdot)$, $\mathbf{e}_A$, $\mathbf{e}_B$, and $\mathbf{e}_N$ are determined by (19). Tie line information belongs to $\mathbf{h}_N(\cdot)$. In this subsection, we assume that i) true measurements $\mathbf{z} = col(\mathbf{z}_A, \mathbf{z}_B, \mathbf{z}_N)$ can pass IEGS BDD, ii) intruders have complete/zero network information of the attacking/non-attacking region, i.e., they know $\mathbf{h}_A(\cdot)$ and $\mathbf{h}_B^A(\cdot)$ and do not know $\mathbf{h}_B^N(\cdot)$ or $\mathbf{h}_N(\cdot)$. iii) intruders can get access to all/zero measurement in the attacking/non-attacking region, i.e., they know $\mathbf{z}_A$ and $\mathbf{z}_B$ and do not know $\mathbf{z}_N$, iv) intruders have the same estimated states $\hat{\mathbf{x}}_A$ and $\hat{\mathbf{x}}_B$ as those estimated by IEGS operators. The above assumptions are quite similar to the assumptions adopted in FDIAs on power systems [16]-[20].

Then, we introduce the proposed FDIAs on IEGSs with local network information. Let $\Delta\mathbf{z}_A$, $\Delta\mathbf{z}_B$, and $\Delta\mathbf{z}_N$ be the injected false data into measurements $\mathbf{z}_A$, $\mathbf{z}_B$, and $\mathbf{z}_N$, respectively, and let $\Delta\mathbf{x}_A$, $\Delta\mathbf{x}_B$, and $\Delta\mathbf{x}_N$ be the variations in estimated states $\hat{\mathbf{x}}_A$, $\hat{\mathbf{x}}_B$, and $\hat{\mathbf{x}}_N$, after the attack, respectively. If an attack $\Delta\mathbf{z} = col(\Delta\mathbf{z}_A, \Delta\mathbf{z}_B, \Delta\mathbf{z}_N)$ satisfies

$$\Delta\mathbf{z}_A = \mathbf{h}_A(\hat{\mathbf{x}}_A+\Delta\mathbf{x}_A,\hat{\mathbf{x}}_B) - \mathbf{h}_A(\hat{\mathbf{x}}_A,\hat{\mathbf{x}}_B), \qquad (20a)$$

$$\Delta\mathbf{z}_B = \mathbf{h}_B^A(\hat{\mathbf{x}}_A+\Delta\mathbf{x}_A,\hat{\mathbf{x}}_B) - \mathbf{h}_B^A(\hat{\mathbf{x}}_A,\hat{\mathbf{x}}_B), \qquad (20b)$$

$$\Delta\mathbf{z}_N = \mathbf{0}, \qquad (20c)$$

where $\Delta\mathbf{x}_A$ can be any real vector, this attack $\Delta\mathbf{z}$ is an FDIA, i.e., it can compromise an IEGS without being detected by IEGS BDD (9). Note that conditions (20) implicitly require that $\Delta\mathbf{x}_B$ and $\Delta\mathbf{x}_N$ are both zero vectors.

To show that, we first check IEGS BDD (9a). For falsified measurements $\mathbf{z} + \Delta\mathbf{z}$, (9a) becomes

$$\|\mathbf{r}_{\text{bad}}\| = \left\| \begin{bmatrix}\mathbf{z}_A + \Delta\mathbf{z}_A - \mathbf{h}_A(\hat{\mathbf{x}}_A+\Delta\mathbf{x}_A,\hat{\mathbf{x}}_B)\\ \mathbf{z}_B + \Delta\mathbf{z}_B - (\mathbf{h}_B^A(\hat{\mathbf{x}}_A+\Delta\mathbf{x}_A,\hat{\mathbf{x}}_B) + \mathbf{h}_B^N(\hat{\mathbf{x}}_N,\hat{\mathbf{x}}_B))\\ \mathbf{z}_N + \Delta\mathbf{z}_N - \mathbf{h}_N(\hat{\mathbf{x}}_N,\hat{\mathbf{x}}_B)\end{bmatrix} \right\|$$

$$= \left\| \begin{bmatrix}\mathbf{z}_A + \mathbf{h}_A(\hat{\mathbf{x}}_A+\Delta\mathbf{x}_A,\hat{\mathbf{x}}_B) - \mathbf{h}_A(\hat{\mathbf{x}}_A,\hat{\mathbf{x}}_B) - \mathbf{h}_A(\hat{\mathbf{x}}_A+\Delta\mathbf{x}_A,\hat{\mathbf{x}}_B)\\ \mathbf{z}_B + \mathbf{h}_B^A(\hat{\mathbf{x}}_A+\Delta\mathbf{x}_A,\hat{\mathbf{x}}_B) - \mathbf{h}_B^A(\hat{\mathbf{x}}_A,\hat{\mathbf{x}}_B) - (\mathbf{h}_B^A(\hat{\mathbf{x}}_A+\Delta\mathbf{x}_A,\hat{\mathbf{x}}_B) + \mathbf{h}_B^N(\hat{\mathbf{x}}_N,\hat{\mathbf{x}}_B))\\ \mathbf{z}_N + \mathbf{0} - \mathbf{h}_N(\hat{\mathbf{x}}_N,\hat{\mathbf{x}}_B)\end{bmatrix} \right\|$$

$$= \left\| \begin{bmatrix}\mathbf{z}_A - \mathbf{h}_A(\hat{\mathbf{x}}_A,\hat{\mathbf{x}}_B)\\ \mathbf{z}_B - (\mathbf{h}_B^A(\hat{\mathbf{x}}_A,\hat{\mathbf{x}}_B) + \mathbf{h}_B^N(\hat{\mathbf{x}}_N,\hat{\mathbf{x}}_B))\\ \mathbf{z}_N - \mathbf{h}_N(\hat{\mathbf{x}}_N,\hat{\mathbf{x}}_B)\end{bmatrix} \right\| = \|\mathbf{r}\| \le \tau,$$

The equation in the last row is derived from (20), indicating that the FDIA $\Delta\mathbf{z}$ can bypass (9a). Then, we check if this FDIA can be detected by IEGS BDD (9b). According to Definition 4, the power buses and gas nodes connected to gas-fired generators or P2G facilities in an attacking region are boundary nodes. According to (20), FDIA $\Delta\mathbf{z}$ does not alter the states of boundary nodes and thus is undetectable by (9b). Overall, we conclude that FDIA $\Delta\mathbf{z}$ cannot be detected by IEGS BDD (9).

***Remark 5:*** To guarantee the stealthiness of an FDIA, injected false data $\Delta\mathbf{z}$ can be derived by setting $\Delta\mathbf{x}_B = \mathbf{0}$ and $\Delta\mathbf{x}_N = \mathbf{0}$. Theoretically, the variation of estimated states $\hat{\mathbf{x}}_A$, i.e., $\Delta\mathbf{x}_A$, can be any vector. In practice, it can be designed by considering the constraints in Section III.B to enhance stealthiness.

Based on the above analysis, we immediately derive

***Proposition 2:*** Given assumptions i)-iv) (the 3[rd] paragraph in this subsection), intruders can design FDIAs $\Delta\mathbf{z}$ on SE (8) of transmission-level IEGSs with local network (topology and parameter) information (i.e., $\mathbf{h}_A(\cdot)$ and $\mathbf{h}_B^A(\cdot)$) by following conditions (20).

***Remark 6:*** According to Proposition 2, the proposed FDIAs require intruders have accurate estimated states in the attacking region (i.e., $\hat{\mathbf{x}}_A$ and $\hat{\mathbf{x}}_B$). Accurate estimated states are the states estimated by IEGS operators with true measurements $\mathbf{z}$ and are elusive. When intruders do not know $\hat{\mathbf{x}}_A$ or $\hat{\mathbf{x}}_B$, they can still estimate the states in the attacking region (denoted as $\hat{\mathbf{x}}_A'$ and $\hat{\mathbf{x}}_B'$) based on measurements $\mathbf{z}_A$ and $\mathbf{z}_B'$. We take the IEGS in Fig. 1 as an example for clarifying how to generate $\mathbf{z}_B'$ and derive estimated states. Let $p_i' = p_i - p_{ij}$, $q_i' = q_i - q_{ij}$, $g_m' = g_m - g_{mn}$, and $g_u' = g_u - c_{uv}$. Measurements $p_i'$, $q_i'$, $g_m'$, and $g_u'$ are revised power and gas injection measurements at boundary nodes. $\mathbf{z}'_{\mathrm{B}} = col(p'_i, q'_i, g'_m, g'_u)$. Intruders can use $\mathbf{z}_{\mathrm{A}}$ and $\mathbf{z}'_{\mathrm{B}}$ to derive $\hat{\mathbf{x}}'_{\mathrm{A}}$ and $\hat{\mathbf{x}}'_{\mathrm{B}}$. Due to the existence of biases between $\hat{\mathbf{x}}_{\mathrm{A}}$, $\hat{\mathbf{x}}_{\mathrm{B}}$ and $\hat{\mathbf{x}}'_{\mathrm{A}}$, $\hat{\mathbf{x}}'_{\mathrm{B}}$, intruders may consider analyzing the impact of biases on the stealthiness of any FDIAs by using the method proposed in Section III.A. For the attacking area in an attacking region, intruders may use the method proposed in [35]. Note that the stealthiness enhancement methods in Section III.B also apply to the FDIAs in this subsection if we consider the power buses connected by detailed compressor models as boundary nodes.

### B. FDIAs on IEGSs With Only Local Topology Information

This subsection further relaxes the assumption in Section IV.A, which requires intruders have local network (topology and parameter) information. Specifically, in this subsection, we assume that i) true measurements $\mathbf{z}$ can pass IEGS BDD, ii) intruders have complete topology information of the attacking region, zero parameter information of the attacking region, and zero network (topology and parameter) information of the non-attacking region, iii) intruders have no access to the measurements in the non-attacking region. Accordingly, we have

***Lemma 1:*** If intruders only have local topology information of an IEGS, the stealthiness of injected false data on any power system measurements in this IEGS cannot be guaranteed.

***Proof:*** We start with a *Simple Case*, where intruders have topology information of an entire IEGS, i.e., the attacking region is the entire IEGS. Let constant matrices $\mathbf{B}^{\mathrm{p}}$ and $\mathbf{B}^{\mathrm{g}}$ be the incidence matrices of power and gas systems in an IEGS, respectively. Specifically, matrix $\mathbf{B}^{\mathrm{p}}/\mathbf{B}^{\mathrm{g}}$ has one row for each power bus/gas node and one column for each power transmission line/gas pipeline (either a gas passive pipeline or a gas compressor). The element in row $i$ and column $j$ in $\mathbf{B}^{\mathrm{p}}/\mathbf{B}^{\mathrm{g}}$ is i) 1 if power transmission line/gas pipeline $j$ is connected to power bus/gas node $i$ with power/gas *outflow* to $i$, ii) -1 if power transmission line/ gas pipeline $j$ is connected to power bus/gas node $i$ with power/ gas *inflow* to $i$, and iii) 0 if they are not connected. Let vector $\mathbf{p}_{IJ} = col(p_{ij})$, $\{i, j\} \in \mathcal{P}_l$. Similarly, we derive vectors $\mathbf{q}_{IJ}$, $\mathbf{p}_I$, $\mathbf{q}_I$, $\mathbf{v}_I$, $\boldsymbol{\theta}_{I'}$, $\mathbf{g}_{IJ}$, $\mathbf{g}_I$, $\boldsymbol{\pi}_I$, and $\mathbf{c}_{IJ}$. Detailed (mathematical) correlation between estimated measurements (i.e., $\hat{\mathbf{p}}_{IJ}$, $\hat{\mathbf{q}}_{IJ}$, $\hat{\mathbf{p}}_I$, $\hat{\mathbf{q}}_I$, $\hat{\mathbf{v}}_I$, $\hat{\boldsymbol{\theta}}_{I'}$, $\hat{\mathbf{g}}_{IJ}$, $\hat{\mathbf{g}}_I$, $\hat{\boldsymbol{\pi}}_I$, $\hat{\mathbf{c}}_{IJ}$) and estimated states (i.e., $\hat{\mathbf{v}}_I$, $\hat{\boldsymbol{\theta}}_I$, $\hat{\boldsymbol{\pi}}_I$, $\hat{\mathbf{c}}_{IJ}$) are shown as follows:

$$\mathbf{h}^{\mathrm{p}}(\hat{\mathbf{x}}^{\mathrm{p}}) = \begin{bmatrix} [\hat{\mathbf{p}}_{IJ}^{\mathrm{T}}, \hat{\mathbf{q}}_{IJ}^{\mathrm{T}}]^{\mathrm{T}} \\ [\hat{\mathbf{p}}_{I}^{\mathrm{T}}, \hat{\mathbf{q}}_{I}^{\mathrm{T}}]^{\mathrm{T}} \\ [\hat{\mathbf{v}}_{I}^{\mathrm{T}}, \hat{\boldsymbol{\theta}}_{I'}^{\mathrm{T}}]^{\mathrm{T}} \end{bmatrix} = \begin{bmatrix} \mathbf{h}_1^{\mathrm{p}}([\hat{\mathbf{v}}_{I}^{\mathrm{T}}, \hat{\boldsymbol{\theta}}_{I}^{\mathrm{T}}]^{\mathrm{T}}) \\ \mathbf{h}_2^{\mathrm{p}}([\hat{\mathbf{v}}_{I}^{\mathrm{T}}, \hat{\boldsymbol{\theta}}_{I}^{\mathrm{T}}]^{\mathrm{T}}) \\ \mathbf{h}_3^{\mathrm{p}}([\hat{\mathbf{v}}_{I}^{\mathrm{T}}, \hat{\boldsymbol{\theta}}_{I}^{\mathrm{T}}]^{\mathrm{T}}) \end{bmatrix} = \begin{bmatrix} \mathbf{h}_1^{\mathrm{p}}([\hat{\mathbf{v}}_{I}^{\mathrm{T}}, \hat{\boldsymbol{\theta}}_{I}^{\mathrm{T}}]^{\mathrm{T}}) \\ \mathbf{D} \cdot \mathbf{h}_1^{\mathrm{p}}([\hat{\mathbf{v}}_{I}^{\mathrm{T}}, \hat{\boldsymbol{\theta}}_{I}^{\mathrm{T}}]^{\mathrm{T}}) \\ \mathbf{I}^{\mathrm{p}} \cdot [\hat{\mathbf{v}}_{I}^{\mathrm{T}}, \hat{\boldsymbol{\theta}}_{I}^{\mathrm{T}}]^{\mathrm{T}} \end{bmatrix}, \tag{21a}$$

$$\mathbf{h}^{\mathrm{g}}(\hat{\mathbf{x}}^{\mathrm{g}}) = \begin{bmatrix} \hat{\mathbf{g}}_{IJ} \\ \hat{\mathbf{g}}_{I} \\ [\hat{\boldsymbol{\pi}}_{I}^{\mathrm{T}}, \hat{\mathbf{c}}_{IJ}^{\mathrm{T}}]^{\mathrm{T}} \end{bmatrix} = \begin{bmatrix} \mathbf{h}_1^{\mathrm{g}}(\hat{\boldsymbol{\pi}}_I) \\ \mathbf{h}_2^{\mathrm{g}}([\hat{\boldsymbol{\pi}}_{I}^{\mathrm{T}}, \hat{\mathbf{c}}_{IJ}^{\mathrm{T}}]^{\mathrm{T}}) \\ \mathbf{h}_3^{\mathrm{g}}([\hat{\boldsymbol{\pi}}_{I}^{\mathrm{T}}, \hat{\mathbf{c}}_{IJ}^{\mathrm{T}}]^{\mathrm{T}}) \end{bmatrix} = \begin{bmatrix} \mathbf{h}_1^{\mathrm{g}}(\hat{\boldsymbol{\pi}}) \\ \mathbf{B}_{\mathrm{h}}^{\mathrm{g}} \cdot \mathbf{h}_1^{\mathrm{g}}(\hat{\boldsymbol{\pi}}) + \mathbf{B}_{\mathrm{c}}^{\mathrm{g}} \cdot \hat{\mathbf{c}}_{IJ} \\ \mathbf{I}^{\mathrm{g}} \cdot [\hat{\boldsymbol{\pi}}_{I}^{\mathrm{T}}, \hat{\mathbf{c}}_{IJ}^{\mathrm{T}}]^{\mathrm{T}} \end{bmatrix}, \tag{21b}$$

$$\begin{aligned} \mathbf{T} \cdot \mathbf{h}_{\mathrm{c}}^{\mathrm{p}}(\hat{\mathbf{x}}^{\mathrm{p}}) &= \mathbf{T}^{\mathrm{p}} \cdot \mathbf{D}' \cdot \mathbf{h}_1^{\mathrm{p}}([\hat{\mathbf{v}}_{I}^{\mathrm{T}}, \hat{\boldsymbol{\theta}}_{I}^{\mathrm{T}}]^{\mathrm{T}}) \\ &= \mathbf{h}_{\mathrm{c}}^{\mathrm{g}}(\hat{\mathbf{x}}^{\mathrm{g}}) = \mathbf{T}^{\mathrm{g}} \cdot \mathbf{B}^{\mathrm{g}} \cdot [\mathbf{h}_1^{\mathrm{g}}(\hat{\boldsymbol{\pi}})^{\mathrm{T}}, \hat{\mathbf{c}}_{IJ}^{\mathrm{T}}]^{\mathrm{T}}. \end{aligned} \tag{21c}$$

$\mathbf{h}_1^{\mathrm{p}}(\cdot)$, $\mathbf{h}_2^{\mathrm{p}}(\cdot)$, $\mathbf{h}_1^{\mathrm{g}}(\cdot)$, and $\mathbf{h}_2^{\mathrm{g}}(\cdot)$ refer to the mappings (A.1)-(A.2), (A.3)-(A.4), (A.5), and (A.6), respectively, in Appendix A. $\mathbf{h}_3^{\mathrm{p}}(\cdot)$ and $\mathbf{h}_3^{\mathrm{g}}(\cdot)$ denote mappings $\mathbf{I}^{\mathrm{p}}$ and $\mathbf{I}^{\mathrm{g}}$, respectively. $\mathbf{I}^{\mathrm{g}}$ is an identity matrix, and $\mathbf{I}^{\mathrm{p}}$, $\mathbf{T}^{\mathrm{p}}$ and $\mathbf{T}^{\mathrm{g}}$ are constant matrices. $\mathbf{B}^{\mathrm{g}} = [\mathbf{B}_{\mathrm{h}}^{\mathrm{g}}\ \mathbf{B}_{\mathrm{c}}^{\mathrm{g}}]$. Constant matrices

$$\mathbf{D} = \begin{bmatrix} \mathbf{B}^{\mathrm{p}} & \mathbf{M}_0 \\ \mathbf{M}_0^{\mathrm{T}} & \mathbf{B}^{\mathrm{p}} \end{bmatrix},\ \mathbf{D}' = \begin{bmatrix} \mathbf{B}^{\mathrm{p}} & \mathbf{M}_0 \\ \mathbf{M}_0^{\mathrm{T}} & \mathbf{N}_0 \end{bmatrix},$$

where $\mathbf{M}_0$ and $\mathbf{N}_0$ are all-zero matrices. Mathematically, the Simple Case means that intruders know matrices $\mathbf{D}$, $\mathbf{D}'$, $\mathbf{I}^{\mathrm{p}}$, $\mathbf{T}^{\mathrm{p}}$, $\mathbf{B}_{\mathrm{h}}^{\mathrm{g}}$, $\mathbf{B}_{\mathrm{c}}^{\mathrm{g}}$, $\mathbf{I}^{\mathrm{g}}$, and $\mathbf{T}^{\mathrm{p}}$ but do not know $\mathbf{h}_1^{\mathrm{p}}(\cdot)$, $\mathbf{h}_2^{\mathrm{p}}(\cdot)$, $\mathbf{h}_{\mathrm{c}}^{\mathrm{p}}(\cdot)$, $\mathbf{h}_1^{\mathrm{g}}(\cdot)$, $\mathbf{h}_2^{\mathrm{g}}(\cdot)$, $\mathbf{h}_{\mathrm{c}}^{\mathrm{g}}(\cdot)$, and $\mathbf{T}$, as intruders have no parameter information.

If intruders have the topology information of an entire IEGS and launch an FDIA on *power system measurements*, according to (21a), except for Special FDIA 1[2], the stealthiness of any FDIA cannot be guaranteed. This is because intruders cannot precisely quantify the injected false data without parameter information (i.e., $\mathbf{h}_1^{\mathrm{p}}(\cdot)$ and $\mathbf{h}_2^{\mathrm{p}}(\cdot)$ in (21a)). Thus, the FDIA has a high possibility of incurring inconsistency between falsified measurements, which are detectable by BDD. For the general case where intruders have *local* topology information of an IEGS, the stealthiness of Special FDIA 1 cannot be guaranteed, either. This is because intruders cannot inject false data into the phase angle measurements in the non-attacking region (according to its definition), where phasor measurement units (PMUs) may be installed. This completes the proof. ■

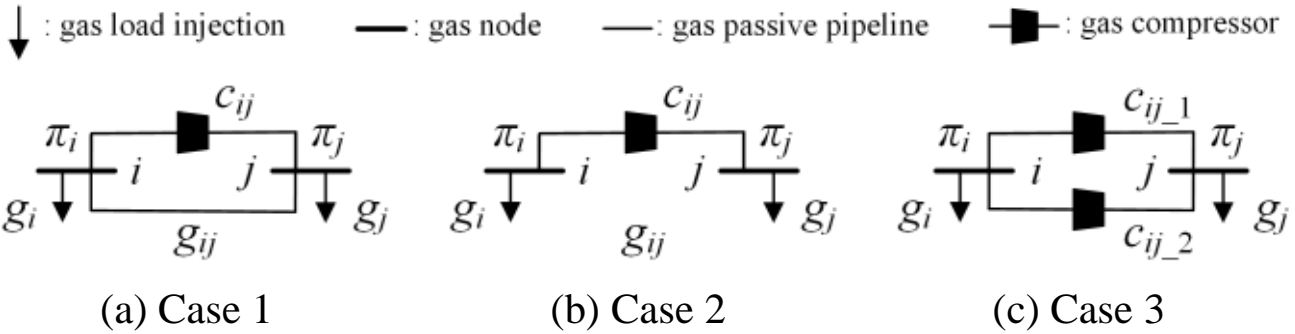

(a) Case 1 (b) Case 2 (c) Case 3

Fig. 2. Illustrative FDIAs on an IEGS with only network topology information.

Lemma 1 indicates that intruders should avoid launching FDIAs on any power subsystem measurements in an IEGS with only local topology information. Counterintuitively, we find that intruders can launch FDIAs on gas subsystem measurements with only local topology information, i.e.,

***Lemma 2:*** Given assumptions i)-iii) (the 1st paragraph in this subsection), intruders can launch FDIAs on SE (8) of a transmission-level IEGS with only local topology information if this local region has nonzero gas load nodes that are connected by at least one gas compressor.

***Proof:*** We proceed with the Simple Case (in the proof of Lemma 1), where intruders have the topology information of an *entire* IEGS and launch an FDIA on *gas subsystem measurements*. The second row of (21b) explicitly show $\mathbf{g}_I$ has an affine relation with $\mathbf{c}_{IJ}$. Intruders know this affine relation, as they know $\mathbf{B}_{\mathrm{h}}^{\mathrm{g}}$ and $\mathbf{B}_{\mathrm{c}}^{\mathrm{g}}$. Theoretically, intruders can launch FDIAs on gas subsystem measurements $\mathbf{g}_I$ and $\mathbf{c}_{IJ}$.

Mathematically, if intruders manage to inject false data $\Delta g_i$ into gas compressor $c_{ij}$, where the connected gas nodes $i$ and $j$ are non-zero gas load nodes, intruders can modify gas injection $g_i$ and $g_j$ by $\mathbf{B}_{\mathrm{c}}^{\mathrm{g}}[i, :]\cdot\Delta\mathbf{c}_{ij}$ and $\mathbf{B}_{\mathrm{c}}^{\mathrm{g}}[j, :]\cdot\Delta\mathbf{c}_{ij}$, respectively. $\mathbf{B}_{\mathrm{c}}^{\mathrm{g}}[i, :]$ is the $i$-th row of matrix $\mathbf{B}_{\mathrm{c}}^{\mathrm{g}}$. Attack vector $\Delta\mathbf{c}_{ij} = col(0, \cdots, 0,$

[2] Special FDIA 1 refers to the FDIA where intruders inject the same false data into all phase angle measurements $\theta_{i'}$, $i' \in \mathcal{P}'_n$. Since the values of trigonometric terms in (A.1) and (A.2) (i.e., cos(·) and sin(·)) remain the same before and after Special FDIA 1, the values of measurements $\mathbf{p}_{IJ}$, $\mathbf{q}_{IJ}$, $\mathbf{p}_I$, $\mathbf{q}_I$, and $\mathbf{v}_I$ in (21a) remain the same. Thus, Special FDIA 1 is stealthy.

$\Delta g_i$, 0, ⋯, 0), where the position of $\Delta g_i$ in $\Delta \mathbf{c}_{ij}$ is the same as that of $\hat{c}_{ij}$ in $\hat{\mathbf{c}}_{IJ}$. Equation (16b) still holds after the modification, i.e., adding the same amount to both left- and right-hand sides of the second row of (21b). Since this attack does not affect other state variables, equations (21a) and (21c) also hold, indicating that this attack is an FDIA. Physically, for Case 1 in Fig. 2, where gas nodes $i$ and $j$ are non-zero gas load nodes, attacks $col(\Delta g_i, -\Delta g_i, \Delta g_i)$ (on measurements $col(g_i, g_j, c_{ij})$) are FDIAs. In fact, the FDIAs, i.e., $col(\Delta g_i, -\Delta g_i, \Delta g_i)$ on measurements $col(g_i, g_j, c_{ij})$, are general FDIAs, as they are still valid if we remove gas passive pipeline $g_{ij}$ (e.g., Case 2) or add gas passive pipelines and/or gas compressors to Case 2 (e.g., Case 3), enabling Case 1 to cover all scenarios in Lemma 2. This completes the proof. ■

Lemma 2 confirms the existence of FDIAs on an IEGS even if intruders only have local topology information of the IEGS. These FDIAs redistribute gas loads at gas nodes $i$ and $j$ (via gas compressor $c_{ij}$) and are termed *gas load redistribution attacks.* In addition to Lemma 2, we have another type of FDIAs on IEGSs with only local topology information. Lemma 3 shows details.

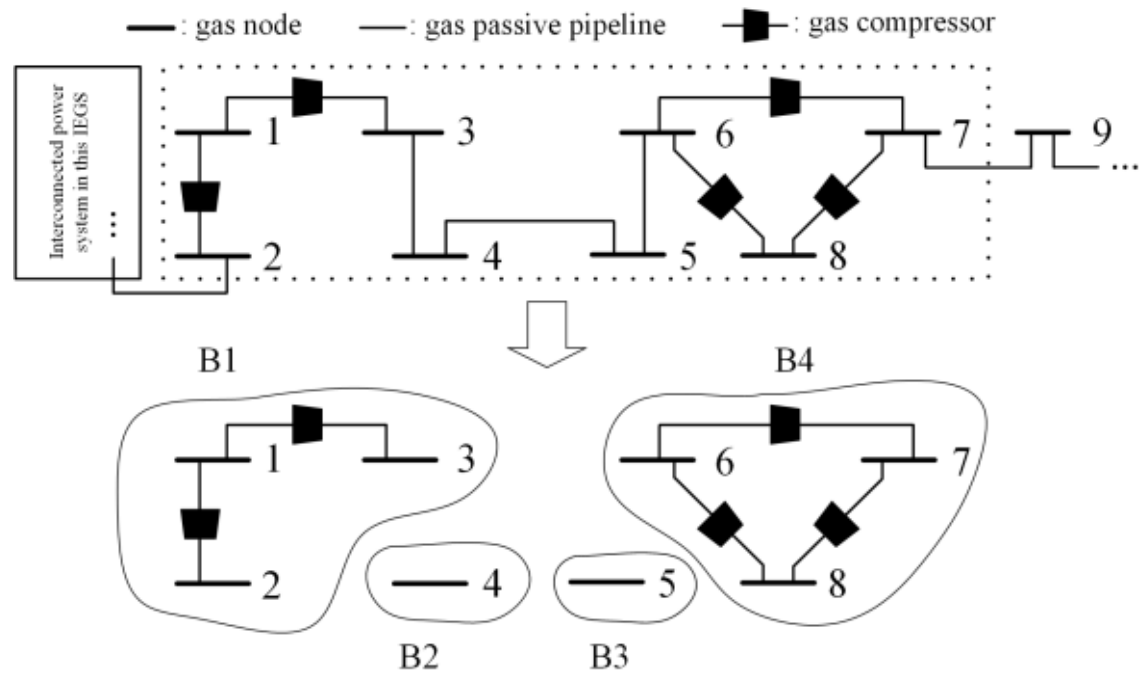

Fig. 3. Illustrative FDIAs on an IEGS with only network topology information of a 8-node gas system containing meshed gas compressors.

***Lemma 3**:* Given assumptions i)-iii) (the 1st paragraph in this subsection), intruders can launch FDIAs on SE (8) of a transmission-level IEGS with only local topology information if this local region has meshed gas compressors.

***Proof**:* Mathematically, if intruders manage to inject false data $\Delta c_{ij}$ and $-\Delta c_{ij}$ into gas compressor $c_{ij_1}$ and $c_{ij_2}$, respectively, where $c_{ij_1}$ and $c_{ij_2}$ connect the same gas nodes $i$ and $j$, we always have $\mathbf{B}_{\mathrm{c}}^{\mathrm{g}} \cdot \Delta \mathbf{c}'_{ij} = \mathbf{0}$, i.e., adding a zero vector to the right hand side of the second row of (21b), so that equation (16b) still holds. The attack vector $\Delta \mathbf{c}'_{ij} = col(0, \cdots, 0, \Delta c_{ij}, 0, \cdots, 0, -\Delta c_{ij}, 0, \cdots, 0)$, and the positions of $\Delta c_{ij}$ and $-\Delta c_{ij}$ in $\Delta \mathbf{c}'_{ij}$ are the same as the positions of $\hat{c}_{ij_1}$ and $\hat{c}_{ij_2}$ in $\hat{\mathbf{c}}_{IJ}$, respectively. Since this attack does not affect other state variables, equations (21a) and (21c) also hold, indicating that this attack is an FDIA. Physically, for Case 3 in Fig. 2, attacks $col(\Delta c_{ij}, -\Delta c_{ij})$ (on measurements $col(c_{ij_1}, c_{ij_2})$) are FDIAs. In fact, Case 3 can be further generalized as meshed gas compressors, where the gas loads in nodes $i$ and $j$ are not required anymore. Fig. 3 gives an example. Intruders are assumed to have local topology information of an IEGS, i.e., the area within the dotted rectangle in Fig. 3 (an eight-node gas system). To see if this area contains meshed gas compressors, we remove gas passive pipelines and only consider gas nodes and gas compressors. Consequently, this 8-node gas system is separated into 4 connected subsystems, B1-B4. Subsystem B4 contains meshed gas compressors. Intruders can launch FDIAs $col(\Delta c_{ij}, \Delta c_{ij}, -\Delta c_{ij})$ (on measurements $col(c_{67}, c_{78}, c_{68})$) on this IEGS. The stealthiness of these FDIAs is similar to the above FDIAs in this proof. In fact, we can always derive FDIAs on B4 (similar to the general FDIAs $col(\Delta c_{ij}, \Delta c_{ij}, -\Delta c_{ij})$) if more gas nodes and gas compressors are added (to B4) for generating a larger mesh (i.e., more meshed compressors), enabling B4 to cover all scenarios in Lemma 3. This completes the proof. ■

The FDIAs in Lemma 3 redistribute gas flows in gas compressors are termed *flow redistribution attacks*. Based on *Lemmas 1-3*, we immediately derive:

***Proposition 3**:* Given assumptions i)-iii) (the 1st paragraph in this subsection), intruders can launch FDIAs on SE (8) of a transmission-level IEGS with only local topology information if this local region has either meshed gas compressors or non-zero gas load nodes that are connected by at least one gas compressor.

***Remark 7**:* Different from Proposition 1 and Proposition 2, Proposition 3 does not require intruders have measurements or the same estimated states $\hat{\mathbf{x}}$ estimated by IEGS operators. From this perspective, the compressors in transmission-level IEGSs are quite vulnerable to FDIAs and need extra protections. Note that the FDIAs shown in this subsection are still effective in compromising the IEGS with detailed gas-driven compressor models when satisfying the conditions in Lemma 2.

***Remark 8**:* By studying intruders' behavior, we gain some insights into designing defense strategies against FDIAs. Specifically, based on Lemma 1, the power system in an IEGS does not need any defense action against FDIAs when intruders have only local network topology information of this IEGS. Differently, the gas system in an IEGS needs extra actions to ensure its cybersecurity. Based on Proposition 3, some special topologies related to gas compressors should be avoided in an IEGS from a cybersecurity perspective, e.g., meshed gas compressors and non-zero gas load nodes that are connected by at least one gas compressor. IEGS operators may take it into consideration when deploying gas compressors.

Before case study, in Table II, we summarize the proposed FDIAs and compare them with the FDIAs in previous works.

## V. Case Study

To validate the effectiveness of FDIAs on IEGSs with complete and incomplete network information, this section conducts tests on an integrated 9-bus-electricity-7-node-gas system (IEGS-9-7) and an integrated 39-bus-electricity-20-node-gas system (IEGS-39-20). Measurements include: i) real and reactive power flows in both ends of power transmission lines, ii) real and reactive power injections at all power buses, iii) voltage magnitudes at all power buses, iv) phase angles at power bus $i'$, $i' \in \mathcal{P}'_n$, v) gas flows in gas passive pipelines and gas compressors, vi) gas injections at all gas nodes, and vii) nodal pressures at all gas nodes. Measurements are generated by adding random noise to optimal energy flow values. *IEGS data and*

*(generated) measurements are listed in [28].* Tests are coded by using Julia 1.6.2 with JuMP on a laptop with an Intel(R) i58265U CPU. The solver is Ipopt 3.13.4.

TABLE II
SUMMARY OF THE PROPOSED FDIAS AND COMPARISON

| Intruders' information | FDIAs | Conditions£ | Differences from previous works [5]-[20] |
|---|---|---|---|
| Complete IEGS info.† | Yes* | Proposition 1 | Considering CAI |
| Local IEGS topology and parameter info. | Yes* | Proposition 2 | Not studied before |
| Local IEGS topology info. | No on power subsystem** | NA† | Not studied before |
| | Yes on gas subsystem‡ | Proposition 3 | Not studied before |

*: "Yes" denotes there exist FDIAs on IEGSs with the system information indicated in the first column of this table if the conditions in the third column of this table are satisfied.
**: "No on power subsystem" denotes the stealthiness of injected false data on power system measurements in an IEGS with only local topology information of this IEGS cannot be guaranteed.
‡: "Yes on gas subsystem" denotes there exist FDIAs on the gas system in an IEGS with only local topology information of this IEGS if the conditions in Proposition 3 are satisfied.
†: Info. and NA are short for information and not applicable, respectively.
£: Conditions refer to the conditions under which the proposed FDIA models can generate FDIAs.

### A. Integrated 9-Bus-Electricity-7-Node-Gas System

The IEGS-9-7 contains three generators (coal-fired generator $G_1$ and gas-fired generators $G_2$ and $G_3$), two gas wells ($GW_1$ and $GW_2$), three power loads ($PL_1$, $PL_2$, and $PL_3$), and three gas loads ($GL_1$, $GL_2$, and $GL_3$).

This subsection tests conditions (10) for designing *FDIAs on IEGSs with complete network information* and explains why FDIAs proposed for pure power systems [5]-[7] cannot be successfully launched in the power subsystem of an IEGS. Measurement errors are set to 0 mean and $2e^{-3}$ variance. The attack targets are voltage magnitude $v_2$, and power and gas injections $p_1$ and $g_6$, respectively, in measurement $\mathbf{z}$. Particularly, the proposed FDIAs on IEGSs (S-B1) are compared with the FDIAs designed for pure power systems [6] (S-B2) to compromise the power subsystem in IEGS-9-7.

TABLE III
FDIAS ON IEGSS WITH COMPLETE NETWORK INFORMATION

| Scenario | | S-B1 | S-B2 | S-B3 |
|---|---|---|---|---|
| FDIA model | | Proposed FDIAs for IEGSs | FDIAs for pure power systems in [6] | Proposed FDIAs for IEGSs |
| Attack target | | $v_2$ | $v_2$ | $p_1$ and $g_6$ |
| Affected power state | | $v_2$ | $v_2$ | $v_2, v_4, v_8, v_9, \theta_1, \theta_2, \theta_4, \theta_5, \theta_7, \theta_8, \theta_9$ |
| Affected gas state | | $c_{42}, \pi_1, \pi_4, \pi_7$ | - | $c_{42}, \pi_1, \pi_2, \pi_4, \pi_6, \pi_7$ |
| Residual $\|\|\mathbf{r}\|\|$ | Before FDIA | $1.2757e^{-2}$ | $1.2757e^{-2}$ | $1.2757e^{-2}$ |
| | After FDIA | $1.2754e^{-2}$ | $10.8227e^{-2}$ | $1.2752e^{-2}$ |

Table III shows that FDIAs on the IEGS exist for achieving different attack targets. *See [28] for the complete test data.* Specifically, for S-B1, the FDIA (on voltage measurement $v_2$) induces *fake voltage violation* by shifting measurement $v_2$ from 1.09 p.u. (before the FDIA) to 1.19 p.u. (after the FDIA). (The upper bound is 1.1 p.u.) Note that although the attack target of S-B1 is not a gas system measurement, intruders still need to inject false data into both power and gas system measurements to ensure the stealthiness of this FDIA. However, for S-B2, residual $||\mathbf{r}||$ soars and violates the threshold, since intruders do not consider the gas subsystem and ignore (10b). Consequently, this attack is detectable by IEGS BDD (9). For S-B3, the FDIA (on power injection $p_1$ and gas injection $g_6$) incurs *fake uneconomic dispatch,* as this FDIA alters the measurements of the power generator and gas well at power bus 1 and gas node 6. Overall, this case validates i) conditions (10) are effective in designing FDIAs on IEGSs with complete network information; and ii) FDIAs on pure power systems cannot be applied to compromise even the power subsystem in an IEGS due to the violation of (10b).

TABLE IV
FDIAS ON IEGSS WITH DIFFERENT BDD METHODS

| Attack target | | | $v_2$ | |
|---|---|---|---|---|
| BDD Method | | (9) | LAV-based BDD [36] | $\chi^2$ test [37] |
| Residual $\|\|\mathbf{r}\|\|$ | Before FDIA | $1.2757e^{-2}$ | $1.3927e^{-2}$ | - |
| | After FDIA | $1.2754e^{-2}$ | $1.3919e^{-2}$ | - |
| Acceptability | | - | - | Yes |

Since the proposed FDIAs are designed based on weighted-least square IEGS SE (8) and BDD (9), their applicability to other BDD methods remains unclear. We conducted two tests, i.e., the LAV-based BDD and the $\chi^2$ test, for validating their applicability. Test results are presented in Table IV. For LAV-based BDD, intruders still aim at weighted-least square IEGS SE (8) and BDD (9) and launch the same FDIAs as those in S-B1, while IEGSs adopt the LAV SE method to estimate system states and, accordingly, conduct BDD via checking residuals. Based on the test results, the LAV-based BDD cannot detect the proposed FDIAs. For $\chi^2$ test, we set the significance level $\alpha = 0.05$. Test results indicate that this method cannot identify the FDIAs, either. Different from other cyberattacks, FDIAs are sophisticated and can avoid increasing SE residuals. Thus, residual-based BDD may not be applicable for detecting FDIAs. This case validates the applicability of the proposed FDIAs against the LAV-based BDD and the $\chi^2$ test.

### B. Integrated 39-Bus-Electricity-20-Node-Gas System

The IEGS-39-20 has ten generators (seven coal-fired generators and three gas-fired generators), two gas wells, nineteen power loads, and nine gas loads. Its topology is shown in Fig. 4.

TABLE V
FDIAS ON IEGSS WITH LOCAL NETWORK INFORMATION

| Scenario | | S-C1 | S-C2 |
|---|---|---|---|
| FDIA model | | Proposed FDIAs for IEGSs | Proposed FDIAs for IEGSs |
| Attack target | | $v_{23}$ | $p_{36}$ and $g_{17}$ |
| Affected power state | | $v_{21}$ - $v_{24}$, $v_{35}$, $v_{36}$, $\theta_{21}$ - $\theta_{24}$, $\theta_{35}$, $\theta_{36}$ | $v_{19}$ - $v_{24}$, $v_{33}$ - $v_{36}$, $\theta_{19}$ - $\theta_{24}$, $\theta_{33}$ - $\theta_{36}$ |
| Affected gas state | | - | $c_{910}$, $\pi_9$ - $\pi_{11}$, $\pi_{17}$ - $\pi_{20}$ |
| Residual $\|\|\mathbf{r}\|\|$ | Before FDIA | 1.1904 | 1.1904 |
| | After FDIA | 1.1904 | 1.1904 |

*1) FDIAs on IEGSs with local network information.* This part tests conditions (20) for deriving *FDIAs on IEGSs with local network information*, and test results are presented in Table V. Intruders are assumed to have local network (topology and parameter) information of two areas, Area-1 (power buses 16, 19-24, and 33-36) and Area-2 (gas nodes 9-14 and 17-20), in the IEGS-39-20 and the measurements in Area-1 and Area-2. Measurement errors follow the same distribution as those in S-B1. The attack targets of S-C1 and S-C2 are voltage magnitude $v_{23}$ in Area-1, and power and gas injections $p_{36}$ and $g_{17}$ in Area-1 and Area-2, respectively. According to the test results, the FDIA in S-C1 incurs *fake thermal limit violations* due to reactive power flow increase (from -0.36 p.u. (before the FDIA) to -1.75 p.u. (after the FDIA) in the power transmission line connecting power buses 21 and 22), and the FDIA in S-C2 leads to *fake uneconomic dispatch. See [28] for the complete test data.* Since Area-1 does not have gas-fired generators, the FDIA in S-C1 does not affect gas states. This test validates the effectiveness of conditions (20) in designing FDIAs on IEGSs with incomplete network information.

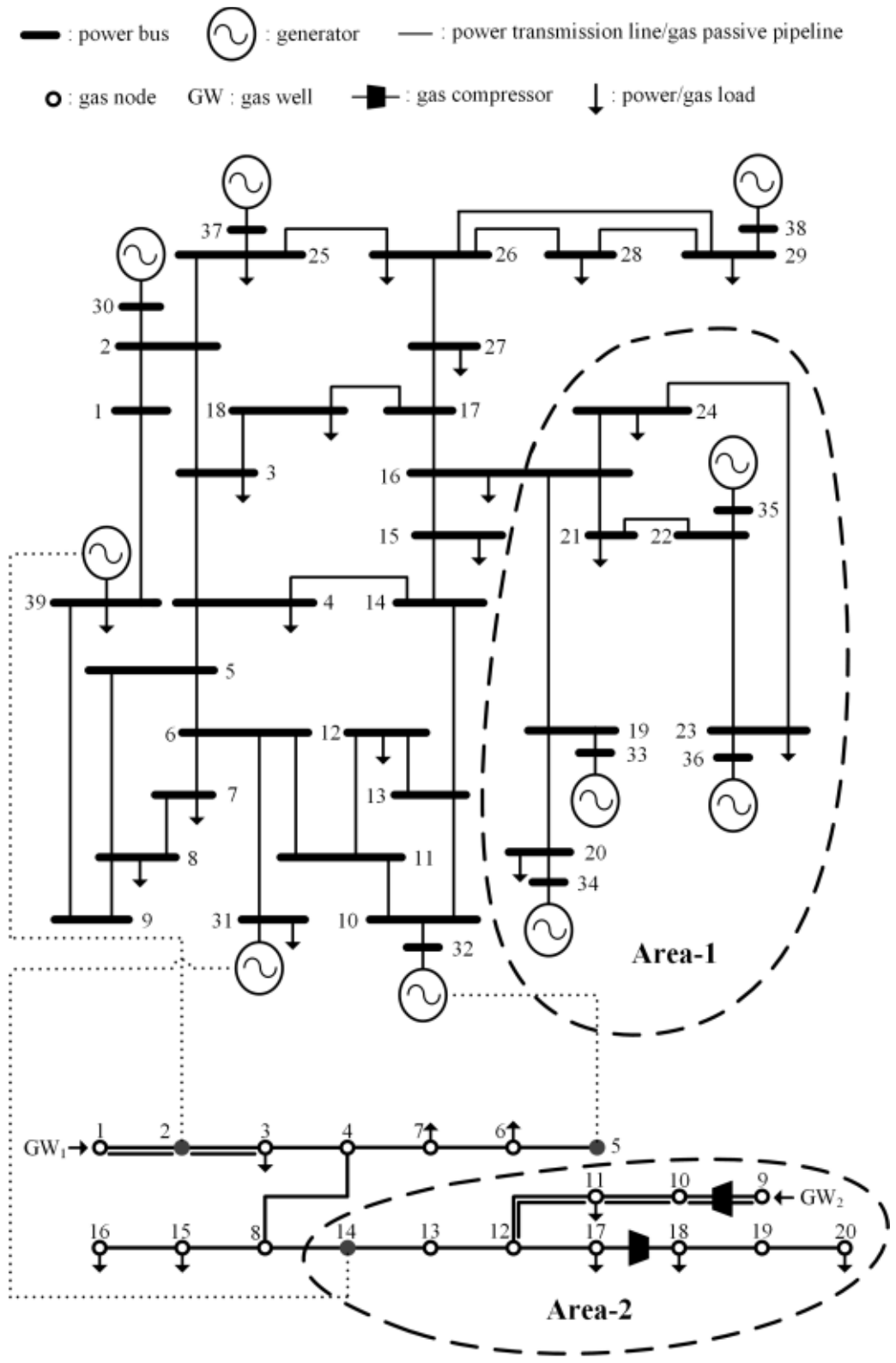

Fig. 4. Topology of IEGS-39-20

*2) FDIAs on IEGSs with only local topology information.* We test Proposition 3 for deriving *FDIAs on IEGSs with only local topology information*, and test results are shown in Table VI. Intruders are assumed to have only topology information of two local areas, i.e., Area-1 (power buses 16, 19-24, and 33-36) and Area-2 (gas nodes 9-14 and 17-20), in the IEGS-39-20. Particularly, in Area-2, gas nodes 9 and 10 are connected by two gas compressors, whose topology is the same as Case 3 in Fig. 2, Gas nodes 17 and 18, each with a gas load, are connected by one gas compressor, whose topology is the same as Case 2 in Fig. 2. As is mentioned in Remark 7, this test does not require that intruders should have the measurements in Area-1 and Area-2. Measurement errors are set to be the same as those in S-C1. The attack targets of S-D1, S-D2, and S-D3 are voltage magnitude $v_{23}$ in Area-1, gas injection $g_{17}$ in Area-2, and gas flows in gas compressors $c'_{9,10}$ and $c''_{9,10}$ in Area-2, respectively. For S-D1, we do not find any FDIA for inducing the voltage violation of $v_{23}$. (The residual A-R is not applicable (NA)). For S-D2, intruders can launch an FDIA for compromising gas injection $g_{17}$. This FDIA is called a *gas load redistribution attack*, as it redistributes gas loads at gas nodes 17 and 18. For S-D3, the FDIA on $c'_{9,10}$ and $c''_{9,10}$ does exist. This FDIA is a *flow redistribution attack*, as it redistributes gas flows in two gas compressors {9, 10}. Note that the proposed FDIAs on IEGSs with only local topology information are general and do not rely on selective regions, provided the conditions in Proposition 3 are satisfied. For example, if intruders aim at compromising compressors $c_{1,2}$ and $c_{2,3}$, they can still derive feasible FDIAs by injecting false data into IEGS measurements $g_1$, $g_3$, $c_{1,2}$, and $c_{2,3}$. *See [28] for the complete test data.* This test numerically validates Lemma 1, Lemma 2, Lemma 3, and Proposition 3 for deriving FDIAs on IEGSs with only local network topology information.

TABLE VI
FDIAS ON IEGSS WITH ONLY LOCAL TOPOLOGY INFORMATION

| Scenario | | S-D1 | S-D2 | S-D3 |
|---|---|---|---|---|
| FDIA model | | Proposed FDIAs for IEGSs | Proposed FDIAs for IEGSs | Proposed FDIAs for IEGSs |
| Attack target | | $v_{23}$ | $g_{17}$ | $c_{9,10}$ |
| Affected power state | | NA | - | - |
| Affected gas state | | NA | $c_{17,18}$ | $c'_{9,10}$, $c''_{9,10}$ |
| Affected IEGS measurements | | NA | $g_{17}$, $g_{18}$, $c_{17,18}$ | $c'_{9,10}$, $c''_{9,10}$ |
| Residual $\|\|\mathbf{r}\|\|$ | Before FDIA | 1.1904 | 1.1904 | 1.1904 |
| | After FDIA | NA | 1.1904 | 1.1904 |

## VI. CONCLUSION

This paper studies FDIAs on IEGSs with complete and incomplete network information. We develop FDIAs on IEGSs with complete network information considering CAI. Then, for the first time, we develop FDIAs on IEGSs with local network (topology and parameter) information and with only local topology information, respectively. For the latter, we mathematically prove and numerically validate that: i) topology-only FDIAs on the power subsystem in an IEGS cannot be devised in general, and ii) topology-only FDIAs on the gas subsystem in an IEGS can be devised, specifically targeting gas compressors. Test results show that: i) intruders should consider CAI when designing FDIAs on IEGSs; ii) IEGSs, especially the gas subsystem in an IEGS, are vulnerable to FDIAs even if intruders do not have complete network information; iii) IEGS operators should pay great attention to the cybersecurity of gas compressors. However, this work still has some limitations: i) the proposed FDIAs rely on IEGS information and measurements (except for the FDIAs in Section IV.B); ii) this work only aims at single-period FDIAs and does not consider energy storage or line packs; iii) this work does not study multi-time scale issues of power and gas systems in an IEGS, iv) this work does not study defense, detection, or mitigation methods against FDIAs. Future works include studying data-free FDIAs, sequential FDIAs on multi-period multi-time scale IEGSs considering IEGS dynamics, energy storage, and line packs, and detection, mitigation, and defense methods against FDIAs.

## APPENDIX

### A. Operation Constraints in IEGSs

In an IEGS, we have the following operation constraints.

$$p_{ij} = v_i^2 \mathrm{G}_{ij} - v_i v_j [\mathrm{G}_{ij} \cos(\theta_i - \theta_j) + \mathrm{B}_{ij} \sin(\theta_i - \theta_j)],\ \forall \{i, j\} \in \mathcal{P}_l, \tag{A.1}$$

$$q_{ij} = \text{-} v_i^2 \mathrm{B}_{ij} - v_i v_j [\mathrm{G}_{ij} \sin(\theta_i - \theta_j) - \mathrm{B}_{ij} \cos(\theta_i - \theta_j)],\ \forall \{i, j\} \in \mathcal{P}_l, \tag{A.2}$$

$$p_i = \sum_{j \in \mathcal{P}_{l(i)}} p_{ij},\ \forall i \in \mathcal{P}_n, \tag{A.3}$$

$$q_i = \sum_{j \in \mathcal{P}_{l(i)}} q_{ij},\ \forall i \in \mathcal{P}_n, \tag{A.4}$$

$$g_{ij} = \sqrt{\mathrm{W}_{ij} (\pi_i^2 - \pi_j^2)},\ \forall \{i, j\} \in \mathcal{G}_l, \tag{A.5}$$

$$g_i = \sum_{j \in \mathcal{G}_{l(i)}} g_{ij} + \sum_{j \in \mathcal{G}_{c(i)}} c_{ij},\ \forall i \in \mathcal{G}_n. \tag{A.6}$$

Constraints (A.1) and (A.2) determine real and reactive power flow in power transmission line $\{i, j\}$, respectively. Constraints (A.3) and (A.4) are real and reactive power balance constraints at power bus $i$, respectively. Set $\mathcal{P}_{l(i)}$ consists of power buses connected to power bus $i$ via power transmission line $l$. Constraint (A.5) is the unidirectional Weymouth equation [8]-[11], [26]. Constraint (A.6) is the gas balance constraints at gas node $i$. Sets $\mathcal{G}_{l(i)}$ and $\mathcal{G}_{c(i)}$ consist of gas nodes connected to gas node $i$ via gas passive pipelines and gas compressors, respectively. Different from AC power flows, where $p_{ij} \neq \text{-}\, p_{ji}$ and $q_{ij} \neq \text{-}\, q_{ji}$, in general, gas flows $g_{ij} = \text{-}\, g_{ji}$ and $c_{ij} = \text{-}\, c_{ji}$ if we do not consider line pack. (The scope of this paper is static SE.) Real and reactive power injections and gas injections are derived by

$$p_i = \sum_{g \in \mathcal{P}_{g(i)} \cup \mathcal{G}_{g(i)}} p_g - \sum_{d \in \mathcal{P}_{d(i)}} \mathrm{P}_d - \sum_{f \in \mathcal{P}_{f(i)}} \gamma_f g_f,\ \forall i \in \mathcal{P}_n, \tag{A.7}$$

$$q_i = \sum_{g \in \mathcal{P}_{g(i)} \cup \mathcal{G}_{g(i)}} q_g - \sum_{d \in \mathcal{P}_{d(i)}} \mathrm{Q}_d,\ \forall i \in \mathcal{P}_n, \tag{A.8}$$

$$g_i = \sum_{w \in \mathcal{G}_{w(i)}} g_w + \sum_{f \in \mathcal{P}_{f(i)}} g_f - \sum_{d \in \mathcal{G}_{d(i)}} \mathrm{G}_d - \sum_{g \in \mathcal{G}_{g(i)}} \gamma_g p_g,\ \forall i \in \mathcal{G}_n. \tag{A.9}$$

Sets $\mathcal{P}_{g(i)}$, $\mathcal{G}_{g(i)}$, $\mathcal{P}_{d(i)}$, and $\mathcal{P}_{f(i)}$ consist of coal-fired power generators, gas-fired power generators, power loads, and P2G facilities connected to power bus $i$, respectively. Sets $\mathcal{G}_{w(i)}$ and $\mathcal{G}_{d(i)}$ consist of gas wells and gas loads connected to gas node $i$, respectively. Particularly, constraint (A.9) reveals power-gas interdependency between the power and gas subsystems in an IEGS (i.e., gas-fired power generators and P2G facilities). Although constraint (A.9) assumes each gas-fired power generator/P2G facilities to be supplied by one gas node/power bus, it can be extended to general cases, i.e., one gas-fired generator/ P2G facilities supplied by multiple gas nodes/power buses.

If there exist (at least) one pair of power bus $i$ and gas node $j$ in an IEGS that satisfies: A.i) power bus $i$ connects only one gas-fired power generator (i.e., without any other component), A.ii) this generator is supplied by gas node $j$, and A.iii) gas node $j$ connects only the gas-fired generator (i.e., without any other component), we have

$$g_j = \gamma_g p_i,\ \forall g \in \mathcal{G}_{g(i)},\ \{i, j\} \in \mathcal{C}_n, \tag{A.10}$$

i.e., one of the coupling constraints in an IEGS, where $\mathcal{C}_n$ is the set of pairs of power buses and gas nodes satisfying A.i)-A.iii).

In addition to gas-fired power generators, P2G facilities in IEGSs, which can transform (surplus) electricity into natural gas, also introduce power-gas interdependency [38]. If there exist (at least) one pair of power bus $i$ and gas node $j$ in an IEGS that satisfies: B.i) gas node $j$ connects only one P2G facility (i.e., without any other component), B.ii) this P2G facility is supplied by power bus $i$, and B.iii) power bus $i$ connects only the P2G facility (i.e., without any other component), we have

$$p_i = \gamma_f g_j,\ \forall f \in \mathcal{P}_{f(i)},\ \{i, j\} \in \mathcal{F}_n, \tag{A.11}$$

i.e., one of the coupling constraints in an IEGS, where $\mathcal{F}_n$ is the set of pairs of power buses and gas nodes satisfying B.i)-B.iii).

### B. Detailed Compressor Model

The formulation of detailed compressor models is presented as follows.

$$\rho_i = \pi_i / (\mathrm{R_s T} u_i), \tag{B.1}$$

$$u_i = 1 + 0.257 (\pi_i / \pi_\mathrm{c}) - 0.533 (\mathrm{T_c / T}) (\pi_i / \pi_\mathrm{c}), \tag{B.2}$$

$$v_{ij} = c_{ij} / \rho_i, \tag{B.3}$$

$$h_{ij} = \mathrm{R_s T} (\kappa / (\kappa - 1)) \left( (\pi_j / \pi_i)^{\kappa - 1/\kappa} - 1 \right) u_i, \tag{B.4}$$

$$p_{c_ij} = c_{ij} h_{ij} / \eta_{ij}, \tag{B.5}$$

$$b_{ij} = f_1 (p_{c_ij}; \mathbf{a}_1), \tag{B.6}$$

$$p_{c_ij} \leq P_{c_ij}, \tag{B.7}$$

$$P_{c_ij} = f_2 (n_{ij}, \mathrm{T_a}; \mathbf{A}_1), \tag{B.8}$$

$$h_{ij} = f_2 (v_{ij}, n_{ij}; \mathbf{A}_2), \tag{B.9}$$

$$\eta_{ij} = f_2 (v_{ij}, n_{ij}; \mathbf{A}_3), \tag{B.10}$$

$$f_1 (v_{ij}; \mathbf{a}_2) \leq h_{ij} \leq f_1 (v_{ij}; \mathbf{a}_3), \tag{B.11}$$

$$\mathrm{N}_{ij}^{\min} \leq n_{ij} \leq \mathrm{N}_{ij}^{\max}, \tag{B.12}$$

$$n_{ij} = v_{ij} / \mathrm{V}_0, \tag{B.13}$$

$$m_{ij} = \left( \mathrm{V}_0 h_{ij} / (2 \pi \eta_{ij}) \right) \rho_i,\ \eta_{ij} = \bar{\eta}_{ij}. \tag{B.14}$$

Table VII lists the definitions of the relevant parameters and variables.

TABLE VII
COMPRESSOR PARAMETERS AND VARIABLES

| Parameter | Description | Parameter | Description |
|---|---|---|---|
| $\mathrm{R_s}$ | Gas constant | $\mathrm{T_a}$ | Ambient temperature |
| T | Temperature | $\mathrm{N}_{ij}^{\mathrm{min/max}}$ | Compressor speed limit |
| $\pi_\mathrm{c}$ | Pseudocritical pressure | $\mathrm{V}_0$ | Operating volume |
| $\mathrm{T_c}$ | Gas temperature | $\bar{\eta}_{ij}$ | Constant adiabatic efficiency |
| $\kappa$ | Isentropic exponent | | |
| **Variable** | **Description** | **Variable** | **Description** |
| $\rho_i$ | Gas density | $\eta_{ij}$ | Adiabatic efficiency |
| $u_i$ | Compressibility factor | $b_{ij}$ | Energy consumption rate |
| $v_{ij}$ | Volumetric gas flow rate | $n_{ij}$ | Compressor speed |
| $h_{ij}$ | Adiabatic enthalpy change | $P_{c_ij}$ | Maximum input power |
| $p_{c_ij}$ | Compressor input power | $m_{ij}$ | Shaft torque |

$\mathbf{a}_1$, $\mathbf{a}_2$, and $\mathbf{a}_3$ are constant vectors, and $\mathbf{A}_1$, $\mathbf{A}_2$, and $\mathbf{A}_3$ are constant matrices. $f_1(x) = \mathrm{a}_{12} x^2 + \mathrm{a}_{11} x + \mathrm{a}_{10}$, and $\mathbf{a}_1 = col(\mathrm{a}_{12}, \mathrm{a}_{11}, \mathrm{a}_{10})$. $f_2(x, y) = col(x^2, x, 1)^\mathrm{T} \mathbf{A}_1 col(x^2, x, 1)$. See Appendix B for the nomenclature about the parameters and variables. Since this paper focuses on single-period FDIAs, we set T, $\mathrm{T_c}$, and $\mathrm{T_a}$ as constants. Constraint (B.1) is the thermodynamical standard equation. Constraint (B.2) relates compressibility factor to gas nodal pressure. Constraint (B.3) transforms mass gas flow into volumetric gas flow. Constraint (B.4) shows the impact of gas nodal pressure and compressibility factor on adiabatic enthalpy. Constraints (B.5) and (B.6) calculate compressor input power and energy consumption rate of the drive of this compressor, respectively. According to [33], the power of a compressor is supplied by its drive, which consumes either electric power (*power-driven compressors*) or gas (*gas-driven compressors*). For power-drive compressors, we assume that their drives are supplied by the power subsystem in an IEGS. For gas-driven compressors, we assume that their drives are supplied by their inflow nodes, respectively, e.g., node $i$ in Fig. 2 is the inflow node of compressor $\{i, j\}$. Constraints (B.7) and (B.8) limit compressor input power. Constraints (B.9)-(B.12) and (B.12)-(B.14) are turbo and piston compressor models, respectively. For a turbo compressor, constraints (B.9) and (B.10) depict compressor speed and adiabatic efficiency, respectively, and constraints (B.11) and (B.12) enforce adiabatic enthalpy change and compressor speed, respectively. For a piston compressor, constraints (B.13) and (B.14) define compressor speed and shaft torque, respectively. The adiabatic efficiency in a piston compressor is a constant [33]. We assume constraint (B.9) has at least one feasible $n_{ij}$ with any $v_{ij}$ and $h_{ij}$.

If we follow Definition 2, the measurements for detailed compressor $ij$, $ij \in \mathcal{G}_c$, are $c_{ij}, \pi_j, \pi_j$. We define state variables for turbo and piston compressors as $\mathbf{x}^{\mathrm{g}}_{\mathrm{T}_ij}$ and $\mathbf{x}^{\mathrm{g}}_{\mathrm{P}_ij}$, respectively. $\mathbf{x}^{\mathrm{g}}_{\mathrm{T}_ij} = col(\rho_i, u_i, v_{ij}, h_{ij}, p_{c_ij}, \eta_{ij}, b_{ij}, n_{ij}, P_{c_ij})$ and $\mathbf{x}^{\mathrm{g}}_{\mathrm{P}_ij} = col(\rho_i, u_i, v_{ij}, h_{ij}, p_{c_ij}, \eta_{ij}, b_{ij}, n_{ij}, m_{ij}, P_{c_ij})$. With measurements $c_{ij}$, $\pi_i, \pi_j$, the values of $\mathbf{x}^{\mathrm{g}}_{\mathrm{T}_ij}$ and $\mathbf{x}^{\mathrm{g}}_{\mathrm{P}_ij}$ are *uniquely* determined by (B.1)-(B.12) and (B.1)-(B.8), (B.12)-(B.14), respectively. Thus, the IEGS with detailed compressor models is *observable*.

The augmented IEGS SE is derived by replacing $\mathbf{h}(\cdot)$, $\mathbf{h}^{\mathrm{p}}_{\mathrm{c}}(\cdot)$, $\mathbf{h}^{\mathrm{g}}_{\mathrm{c}}(\cdot)$, $\mathbf{T}$, $\mathbf{x}$, $\mathbf{x}^{\mathrm{g}}$, and $\hat{\mathbf{x}}$ in IEGS SE (8) with $\underline{\mathbf{h}}(\cdot)$, $\underline{\mathbf{h}}^{\mathrm{p}}_{\mathrm{c}}(\cdot)$, $\underline{\mathbf{h}}^{\mathrm{g}}_{\mathrm{c}}(\cdot)$, $\underline{\mathbf{T}}$, $\underline{\mathbf{x}}$, $\underline{\mathbf{x}}^{\mathrm{g}}$, and $\underline{\hat{\mathbf{x}}}$, respectively. $\underline{\mathbf{h}}(\cdot)$ is derived by replacing (A.7) and (A.9) in $\mathbf{h}(\cdot)$ with (B.15) and (B.16). $\underline{\mathbf{h}}^{\mathrm{p}}_{\mathrm{c}}(\cdot)$, $\underline{\mathbf{h}}^{\mathrm{g}}_{\mathrm{c}}(\cdot)$ and $\underline{\mathbf{T}}$ are derived by adding constraints (B.17) and (B.18) to (8b). $\underline{\mathbf{x}} = col(\mathbf{x}^{\mathrm{p}}, \underline{\mathbf{x}}^{\mathrm{g}})$, $\underline{\mathbf{x}}^{\mathrm{g}} = col(\mathbf{x}^{\mathrm{g}}, \mathbf{x}^{\mathrm{g}}_{\mathrm{T}_ij}, \mathbf{x}^{\mathrm{g}}_{\mathrm{P}_mn})$, $\{i, j\} \in \mathcal{G}^{\mathrm{T}}_c$, $\{m, n\} \in \mathcal{G}^{\mathrm{P}}_c$. $\mathcal{G}^{\mathrm{T}}_c$ and $\mathcal{G}^{\mathrm{P}}_c$ are the sets of turbine and piston compressors, respectively ($\mathcal{G}_c = \mathcal{G}^{\mathrm{T}}_c \cup \mathcal{G}^{\mathrm{P}}_c$). $\underline{\hat{\mathbf{x}}} = col(\hat{\mathbf{x}}^{\mathrm{p}}, \underline{\hat{\mathbf{x}}}^{\mathrm{g}})$ is the estimated IEGS states by the augmented IEGS SE. Accordingly, we have an augmented IEGS BDD for detecting bad data during the augmented IEGS SE, where $\mathbf{h}(\cdot)$, $\mathbf{h}^{\mathrm{p}}_{\mathrm{c}}(\cdot)$, $\mathbf{h}^{\mathrm{g}}_{\mathrm{c}}(\cdot)$, $\mathbf{T}$, and $\hat{\mathbf{x}}$ in IEGS BDD (9) are replaced by $\underline{\mathbf{h}}(\cdot)$, $\underline{\mathbf{h}}^{\mathrm{p}}_{\mathrm{c}}(\cdot)$, $\underline{\mathbf{h}}^{\mathrm{g}}_{\mathrm{c}}(\cdot)$, $\underline{\mathbf{T}}$, and $\underline{\hat{\mathbf{x}}}$, respectively. As is mentioned in Section III.B, based on the augmented Proposition 1, intruders can design FDIAs on the augmented IEGS SE considering detailed compressor models, where $\Delta\underline{\mathbf{x}} = col(\Delta\mathbf{x}^{\mathrm{p}}, \Delta\underline{\mathbf{x}}^{\mathrm{g}})$. $\Delta\underline{\mathbf{x}}^{\mathrm{g}}$ is the variations of estimated states $\underline{\hat{\mathbf{x}}}^{\mathrm{g}}$ after FDIAs.

$$p_i = \sum_{g \in \mathcal{P}_{g(i)} \cup \mathcal{G}_{g(i)}} p_g - \sum_{d \in \mathcal{P}_{d(i)}} \mathrm{P}_d - \sum_{f \in \mathcal{P}_{f(i)}} \gamma_f g_f - \sum_{\{m,\, n\} \in \mathcal{G}^{\mathrm{E}}_{c(i)}} b_{mn}, \ \forall i \in \mathcal{P}_n, \tag{B.15}$$

$$g_i = \sum_{w \in \mathcal{G}_{w(i)}} g_w + \sum_{f \in \mathcal{P}_{f(i)}} g_f - \sum_{d \in \mathcal{G}_{d(i)}} \mathrm{G}_d - \sum_{g \in \mathcal{G}_{g(i)}} \gamma_g p_g - \sum_{\{i,\, j\} \in \mathcal{G}^{\mathrm{G}}_{c(i)}} \gamma_{ij} b_{ij}, \ \forall i \in \mathcal{G}_n, \tag{B.16}$$

$$p_i = b_{mn}, \ \forall \{m, n\} \in \mathcal{G}^{\mathrm{E}}_{c(i)}, i \in \mathcal{D}^{\mathrm{E}}_n, \tag{B.17}$$

$$g_i = \gamma_{ij} b_{ij}, \ \forall \{i, j\} \in \mathcal{G}^{\mathrm{G}}_{c(i)}, i \in \mathcal{D}^{\mathrm{G}}_n. \tag{B.18}$$

Equations (B.15) and (B.16) derive real power injection and gas injection, respectively. Equations (B.17) and (B.18) are coupling constraints, indicating the electricity and gas supplied to power- and gas-driven compressors, respectively. Equation (B.17) holds if there exist (at least) one power bus $i$ in an IEGS satisfying: C.i) power bus $i$ connects only one (power-driven) compressor (i.e., without any other component), and C.ii) the (power-driven) compressor is supplied by power bus $i$. Similarly, Equation (B.18) holds if there exist (at least) one gas node $i$ in an IEGS satisfying: D.i) gas node $i$ connects only one (gas-driven) compressor (i.e., without any other component) and is an inflow node, and D.ii) this (gas-driven) compressor is supplied by (inflow) gas node $i$. Sets $\mathcal{G}^{\mathrm{E}}_{c(i)}$ and $\mathcal{G}^{\mathrm{G}}_{c(i)}$ is composed of power-driven and gas-driven compresssors connected to power bus $i$ and gas (inflow) node $i$, respectively. Sets $\mathcal{D}^{\mathrm{E}}_n$ and $\mathcal{D}^{\mathrm{G}}_n$ consist of the power buses and gas nodes satisfying C.i)-C.ii) and D.i)-D.ii), respectively. $\gamma_{ij}$, $\{i, j\} \in \mathrm{G}^{\mathrm{G}}_{\mathrm{c(i)}}$, is the electricity-gas conversion ratio of gas-driven compressor $\{i, j\}$.

## REFERENCES

[1] U.S. Energy Information Administration (EIA), Annual Energy Review, 2021 [Online]. Available: https://www.eia.gov/totalenergy/data/annual/

[2] International Energy Agency (IEA), Power Systems in Transition: Challenges and Opportunities Ahead for Electricity Security, 2020 [Online]. Available: https://www.iea.org/reports/power-systems-intransition/ cyberresilience

[3] U.S. Department of Energy (DoE), Colonial Pipeline Cyber Incident, 2021 [Online]. Available: https://www.energy.gov/ceser/colonial- pipeline-cyber-incident

[4] United States Government Accountability Office (GAO), Electricity Grid Cybersecurity: DOE Needs to Ensure Its Plans Fully Address Risks to Distribution Systems, 2021, [Online]. Available: https://www.gao.gov/products/gao-21-81

[5] Y. Liu, P. Ning, and M. K. Reiter, "False data injection attacks against state estimation in electric power grids," *ACM Trans. Inf. Syst. Security*, vol. 14, no. 1, May 2011, Art. no. 13.

[6] G. Hug and J. A. Giampapa, "Vulnerability assessment of AC state estimation with respect to false data injection cyber-attacks," *IEEE Trans. Smart Grid*, vol. 3, no. 3, pp. 1362-1370, Sept. 2012.

[7] P. Zhuang and H. Liang, "False data injection attacks against state-of-charge estimation of battery energy storage systems in smart distribution networks," *IEEE Trans. Smart Grid*, vol. 12, no. 3, pp. 2566-2577, May 2021.

[8] B. Zhao, A. J. Lamadrid, R. S. Blum, and S. Kishore, "A coordinated scheme of electricity-gas systems and impacts of a gas system FDI attacks on electricity system," *Int. J. Electr. Power Energy Syst.*, vol. 131, p. 107060, Oct. 2021.

[9] B. Zhao, A. J. Lamadrid, R. S. Blum, and S. Kishore, "A trilevel model against false gas-supply information attacks in electricity systems," *Elect. Power Syst. Res.*, vol. 189, p. 106541, 2020.

[10] I. L. Carreño, A. Scaglione, A. Zlotnik, D. Deka, and K. Sundar, "An adversarial model for attack vector vulnerability analysis on power and gas delivery operations," *Elect. Power Syst. Res.*, vol. 189, p. 106777, 2020.

[11] M. Zadsar, A. Abazari, A. Ameli, J. Yan, and M. Ghafouri, "Prevention and detection of coordinated false data injection attacks on integrated power and gas systems," *IEEE Trans. Power Syst.*, vol. 38, no. 5, pp. 4252-4268, Sept. 2023.

[12] P. Zhao *et al.*, "Cyber-resilient multi-energy management for complex systems," *IEEE Trans. Ind. Informat.*, vol. 18, no. 3, pp. 2144-2159, March 2022.

[13] R.-P. Liu, X. Wang, B. Zeng, and R. Zgheib, "Modeling load redistribution attacks in integrated electricity-gas systems," *IEEE Trans. Smart Grid*, 2024 (early access).

[14] P. Zhao, C. Gu, and D. Huo, "Coordinated risk mitigation strategy for integrated energy systems under cyber-attacks," *IEEE Trans. Power Syst.*, vol. 35, no. 5, pp. 4014-4025, Sept. 2020.

[15] P. Zhao *et al.*, "A cyber-secured operation for water-energy nexus," *IEEE Trans. Power Syst.*, vol. 36, no. 4, pp. 3105-3117, July 2021.

[16] X. Liu and Z. Li, "Local load redistribution attacks in power systems with incomplete network information," *IEEE Trans. Smart Grid*, vol. 5, no. 4, pp. 1665-1676, July 2014.

[17] X. Liu and Z. Li, “False data attacks against AC state estimation with incomplete network information,” *IEEE Trans. Smart Grid*, vol. 8, no. 5, pp. 2239-2248, Sept. 2017.

[18] R. Deng, P. Zhuang, and H. Liang, “False data injection attacks against state estimation in power distribution systems,” *IEEE Trans. Smart Grid*, vol. 10, no. 3, pp. 2871-2881, May 2019.

[19] S. Bi and Y. J. Zhang, “Using covert topological information for defense against malicious attacks on DC state estimation,” *IEEE J. Sel. Areas Commun.*, vol. 32, no. 7, pp. 1471-1485, July 2014.

[20] W.-L. Chin, C.-H. Lee, and T. Jiang, “Blind false data attacks against AC state estimation based on geometric approach in smart grid communications,” *IEEE Trans. Smart Grid*, vol. 9, no. 6, pp. 6298-6306, Nov. 2018.

[21] Y. Song, X. Liu, Z. Li, M. Shahidehpour, and Z. Li, “Intelligent data attacks against power systems using incomplete network information: a review,” *J. Mod. Power Syst. Clean Energy*, vol. 6, no. 4, pp. 630-641, July 2018.

[22] The European Network of Transmission System Operators for Gas (ENT SOG), System Capacity Map [Online]. Available: https://www.entsog.eu/maps

[23] H. Zang, M. Geng, M. Huang, Z. Wei, S. Chen, and G. Sun, “Asynchronous and adaptive state estimation of integrated electricity–gas energy systems,” *IEEE Internet Things J.*, vol. 10, no. 9, pp. 7636-7644, May, 2023.

[24] National Renewable Energy Lab (NREL), Electric Power Grid and Natural Gas Network Operations and Coordination, 2021 [Online]. Available: https://www.nrel.gov/docs/fy20osti/77096.pdf.

[25] T. W. K. Mak, P. V. Hentenryck, A. Zlotnik, and R. Bent, “Dynamic compressor optimization in natural gas pipeline systems,” *Informs J. Comput.*, vol. 31, no. 1, pp. 40–65, Jan. 2019.

[26] Z. Wang and R. S. Blum, “Elimination of undetectable attacks on natural gas networks,” *IEEE Signal Process. Lett.*, vol. 28, pp. 1002-1005, 2021.

[27] J. Zhao and Y. Guo, “State estimation for integrated energy systems: Motivations, advances, and challenges,” *IEEE Task Force*, Tech. Rep. PES-TR118, Dec. 2023.

[28] [Online]. Available: https://sites.google.com/site/rongpengliu1991/home/data/modeling-false-data-injection-attacks-on-integrated-electricity-gas-systems?authuser=0

[29] W. Zheng, W. Wu, A. Gomez-Exposito, B. Zhang, and Y. Guo, “Distributed robust bilinear state estimation for power systems with nonlinear measurements,” *IEEE Trans. Power Syst.*, vol. 32, no. 1, pp. 499-509, Jan. 2017.

[30] Department of Homeland Security, Insider Threat to Utilities [Online]. Available: https://info.publicintelligence.net/DHS-InsiderThreat.pdf.

[31] J. Zhao, L. Mili, and M. Wang, “A generalized false data injection attacks against power system nonlinear state estimator and countermeasures,” *IEEE Trans. Power Syst.*, vol. 33, no. 5, pp. 4868-4877, Sep. 2018.

[32] X. Liu, Y. Song, and Z. Li, “Dummy data attacks in power systems,” *IEEE Trans. Smart Grid*, vol. 11, no. 2, pp. 1792-1795, March 2020.

[33] D. Rose, M. Schmidt, M. C. Steinbach, and B. M. Willert, “Computational optimization of gas compressor stations: MINLP models versus continuous reformulations,” *Math. Methods Oper. Res.*, vol. 83, pp. 409-444, 2016.

[34] Cyber Operations Tracker, “Compromise of a power grid in eastern Ukraine,” 2015 [Online]. Available: https://www.cfr.org/cyber-operations/compromise-power-grid-eastern-ukraine

[35] X. Liu, Z. Bao, D. Lu, and Z. Li, “Modeling of local false data injection attacks with reduced network information,” *IEEE Trans. Smart Grid*, vol. 6, no. 4, pp. 1686-1696, July 2015.

[36] M. Göl and A. Abur, “LAV based robust state estimation for systems measured by PMUs,” *IEEE Trans. Smart Grid*, vol. 5, no. 4, pp. 1808-1814, July 2014.

[37] P. Hu, W. Gao, Y. Li, F. Hua, L. Qiao, and G. Zhang, “Detection of false data injection attacks in smart grid based on joint dynamic and static state estimation,” *IEEE Access*, vol. 11, pp. 45028-45038, 2023.

[38] L. Yang, Y. Xu, W. Gu, and H. Sun, “Distributionally robust chance-constrained optimal power-gas flow under bidirectional interactions considering uncertain wind power,” *IEEE Trans. Smart Grid*, vol. 12, no. 2, pp. 1722-1735, March 2021.